\documentclass{aa}
\usepackage{graphicx}
\begin{document}
\title{The VIMOS-VLT Deep Survey
         \thanks{based on data
         obtained with the European Southern Observatory Very Large
         Telescope, Paranal, Chile, program 070.A-9007(A), and on data
         obtained at the Canada-France-Hawaii Telescope, operated by
         the CNRS of France, CNRC in Canada and the University of Hawaii}
}
\subtitle{Evolution of the galaxy luminosity function up to z~=~2 in first
  epoch data }

\author{
     O.~Ilbert\inst{1}
\and L.~Tresse\inst{1}
\and E.~Zucca\inst{2}
\and S.~Bardelli\inst{2}
\and S.~Arnouts\inst{1}
\and G.~Zamorani\inst{2} 
\and L.~Pozzetti\inst{2} 
\and D.~Bottini\inst{3}
\and B.~Garilli\inst{3}
\and V.~Le~Brun\inst{1}
\and O.~Le~F\`evre\inst{1}
\and D.~Maccagni\inst{3}
\and J.-P.~Picat\inst{4}
\and R. Scaramella\inst{5}
\and M.~Scodeggio\inst{3}
\and G.~Vettolani\inst{5}
\and A.~Zanichelli\inst{5}
\and C.~Adami\inst{1}
\and M.~Arnaboldi\inst{6}
\and M.~Bolzonella \inst{7} 
\and A.~Cappi\inst{2}
\and S.~Charlot\inst{8,9}
\and T.~Contini\inst{4}
\and S.~Foucaud\inst{3}
\and P.~Franzetti\inst{3}
\and I.~Gavignaud \inst{4,12}
\and L.~Guzzo\inst{10}
\and A.~Iovino\inst{10}
\and H.J.~McCracken\inst{9,11}
\and B.~Marano\inst{7}  
\and C.~Marinoni\inst{1}
\and G.~Mathez\inst{4}
\and A.~Mazure\inst{1}
\and B.~Meneux\inst{1}
\and R.~Merighi\inst{2} 
\and S.~Paltani\inst{1}
\and R.~Pello\inst{4}
\and A.~Pollo\inst{10}
\and M.~Radovich\inst{6}
\and M.~Bondi\inst{5}
\and A.~Bongiorno\inst{7}
\and G.~Busarello\inst{6}
\and P.~Ciliegi\inst{2}  
\and Y.~Mellier\inst{9,11}
\and P.~Merluzzi\inst{6}
\and V.~Ripepi\inst{6}
\and D.~Rizzo\inst{4}
}

\offprints{O.~Ilbert, e-mail: Olivier.Ilbert@oamp.fr}
   
\institute{ 
Laboratoire d'Astrophysique de Marseille (UMR 6110), CNRS-Universit\'e de Provence, B.P.8, 13376 Marseille C\'edex 12, France 
\and INAF-Osservatorio Astronomico di Bologna, via Ranzani 1, 40127 Bologna, Italy 
\and INAF-IASF, via Bassini 15, 20133 Milano, Italy  
\and Laboratoire d'Astrophysique de l'Observatoire Midi-Pyr\'en\'ees (UMR 5572), CNRS-Universit\'e Paul Sabatier, 14 avenue E. Belin, 31400 Toulouse, France 
\and INAF-IRA, via Gobetti 101, 40129 Bologna, Italy 
\and INAF-Osservatorio Astronomico di Capodimonte, via Moiariello 16, 80131 Napoli, Italy 
\and Universit\`a di Bologna, Dipartimento di Astronomia, via Ranzani 1, 40127 Bologna, Italy 
\and Max-Planck-Institut f\"ur Astrophysik, Karl-Schwarzschild-Str. 1, 85740 Garching bei M\"unchen, Germany 
\and Institut d'Astrophysique de Paris (UMR 7095), 98 bis Boulevard Arago, 75014 Paris, France 
\and INAF-Osservatorio Astronomico di Brera, via Brera 28, 20121 Milano, Italy 
\and Observatoire de Paris-LERMA, 61 avenue de l'Observatoire, 75014 Paris, France 
\and European Southern Observatory, Karl-Schwarzschild-Str. 2, 85748 Garching bei M\"unchen, Germany 
}

\date{Received September 7, 2004 / Accepted January 31, 2005 }

\abstract{We investigate the evolution of the galaxy luminosity
  function from the VIMOS-VLT Deep Survey (VVDS) from the present to
  $z~=~2$ in five ($U$, $B$, $V$, $R$ and $I$) rest-frame band-passes.
  We use the first epoch VVDS deep sample of 11,034 spectra selected
  at $17.5 \leq I_{AB} \leq 24.0$, on which we apply the Algorithm for
  Luminosity Function (ALF), described in this paper.  We observe a
  substantial evolution with redshift of the global luminosity
  functions in all bands. From $z~=~0.05$ to $z~=~2$, we measure a
  brightening of the characteristic magnitude M$^{*}$ included in the
  magnitude range $1.8-2.5$, $1.7-2.4$, $1.2-1.9$, $1.1-1.8$ and
  $1.0-1.6$ in the $U$, $B$, $V$, $R$ and $I$ rest-frame bands,
  respectively. We confirm this differential evolution of the
  luminosity function with rest-frame wavelength from the measurement
  of the comoving density of bright galaxies ($M \le M^{*}(z~=~0.1)$).
  This density increases by a factor of around $2.6, 2.2, 1.8, 1.5,
  1.5$ between $z=0.05$ and $z=1$ in the $U$, $B$, $V$, $R$, $I$
  bands, respectively. We also measure a possible steepening of the
  faint-end slope of the luminosity functions, with
  $\Delta\alpha~\sim~-0.3$ between $z=0.05$ and $z=1$, similar in all
  bands.

    }

\titlerunning{Evolution of the galaxy global VVDS LF up to $z~=~2$}
\authorrunning{Ilbert O. et al.}
\maketitle

\section{Introduction}

The luminosity function (LF) of field galaxies is a fundamental
diagnostic of the physical processes that act in the formation and
evolution of galaxies. The LF evolution is mainly determined by the
combination of the star formation history in each galaxy and the
gravitational growth of structures, through merging. These two
different processes are better probed by the luminosity emitted in the
blue and red rest-frame wavelengths, respectively. The relative
contribution of these processes to the cosmic history is reflected in
the LF evolution, which therefore is expected to be different as a
function of rest-frame wavelength. Large deep redshift surveys,
combined with multi-color imaging, are necessary to perform this
measurement.

The local LF is now well constrained by the results of two large
spectroscopic surveys: the Two-Degree Field Redshift Survey (2dFGRS;
\cite{Norberg02}2002) and the Sloan Digital Sky Survey (SDSS;
\cite{Blanton03}2003). These measurements of the local LF
represent the local benchmark for all studies of the LF evolution. Up
to $z \simeq 1$ the Canada-France Redshift Survey (CFRS;
\cite{Lilly95}1995) represents a sample of 591 spectroscopic redshifts
of galaxies, from which it was demonstrated that the global LF evolves
with cosmic time.  \cite{Lilly95}(1995) showed that the evolution of the
LF depends on the galaxy population studied. The LF of the red
population shows few changes over the redshift range $0.05 \le z
\le 1$, while the LF of the blue population brightens by about one
magnitude over the same redshift interval. Up to $z \simeq 0.6$, the
Canadian Network for Observational Cosmology Field Galaxy Redshift
Survey (CNOC2; \cite{Lin99}) and the ESO-Sculptor Survey (ESS;
\cite{DeLapparent03}) derived the LFs per spectral
type with spectroscopic redshift samples of $\sim 2000$ and 617
galaxies, respectively. They confirmed a steep faint-end slope of the LF
for the blue galaxy types. At higher redshift, LF measurements based
on photometric redshifts have been derived by, e.g., \cite{Wolf03}(2003)
up to $z<1.2$, \cite{Gabasch04}(2004) up to $z < 5$. Samples of
Lyman-break selected galaxies have also been used to measure the LF at
such high redshift $3 < z < 5$ (e.g., \cite{Steidel99}1999).

The VIMOS (VIsible Multi-Object Spectrograph) VLT (Very Large
Telescope) Deep Survey (VVDS) is a deep spectroscopic survey conducted
over a large area associated with multi-color photometric data
(\cite{LeFevre04_I}). Because of its characteristics, the VVDS is very
well suited for detailed studies of the LF
evolution :
\begin{itemize}
\item the Universe is surveyed over more than 90\% of its current
  age with spectroscopic redshifts, which allows us to measure the LF
  evolution in a coherent way within a single sample from $z = 0.05$
  up to high redshift;
\item the spectroscopic targets are selected on the basis
  of a simple magnitude limit criterion, with no attempt to exclude
  stars or AGNs, or to select objects on the basis of their
  colors or morphology. Therefore, this selection minimizes any
  bias in sampling the galaxy population up to high redshift;
\item the multi-color coverage allows us to span a large
  range of rest-frame wavelengths, related to different physical
  processes, and thus to derive the LFs in several rest-frame bands;
\item the spectra can also be used to derive the LFs as a
  function of specific spectral properties.
\end{itemize}

In this paper, we focus on the deep fields of the VVDS, 
which are the VVDS-0226-04 field (\cite{LeFevre05}2005) and the
VVDS-Chandra Deep Field South (CDFS; \cite{LeFevre04_II}2004b). The
first epoch VVDS deep sample contains 11,034 spectra of objects
selected at $17.5 \le I_{AB}\le 24.0$.  The first goal of the LF
analysis is to characterize the statistical properties of the whole
population. It is the first measurement that a theory of galaxy
formation must account for, because it is free of any of the
possible biases which can arise when the sample is split
into different sub-samples according to various selection criteria
(e.g., rest-frame colors, morphology, spectral properties). In this
paper, we present the measurement of the evolution of the global LF up
to $z = 2$. To investigate the dependence of this
evolution on the rest-frame wavelength, the global LFs are estimated
in five rest-frame bands, which span the wavelength range 3600\AA$\le
\lambda_{eff} \le$ 7840\AA .

The paper is organized as follows. In Section~2 we briefly present the
VVDS Deep first epoch sample. In Section~3 we describe the target
sampling rate and the spectroscopic success rate of our data. In
Section~4 we discuss two points relevant to the estimate of the global
LF with VVDS data. This estimate is performed with our LF
tool named Algorithm for Luminosity Function (ALF), extensively
described in the Appendix.  In Section~5 we present our results,
compared with other literature measurements in Section~6.  Conclusions
are presented in Section~7.  This paper will be followed by an
analysis of the evolution of the LF per spectral type
(\cite{Zucca05}2005), and as a function of environment
(\cite{Ilbert05}2005).

We use a flat lambda ($\Omega_m~=~0.3$, $\Omega_\Lambda~=~0.7$)
cosmology with $h~=~H_{\rm0}/100$~km~s$^{-1}$~Mpc$^{-1}$. Magnitudes
are given in the $AB$ system.

\section{Data description}

We consider the deep spectroscopic sample of the first
epoch data in the VVDS-0226-04 and VVDS-CDFS fields.

McCracken et al. (2003) describe in detail the photometry and
astrometry of the VVDS-0226-04 field acquired with the wide-field 12K
mosaic camera at the Canada-France-Hawaii Telescope (CFHT). The deep
field covers 1.2~deg$^2$ and reaches the limiting
magnitudes of $B_{AB} \sim 26.5$, $V_{AB} \sim 26.2$, $R_{AB} \sim
25.9$ and $I_{AB} \sim 25.0$, corresponding to 50\% completeness.
These data are complete and free of surface brightness selection
effects at $I_{AB} \le 24.0$, corresponding to the limit of the VVDS
spectroscopic sample. Apparent magnitudes are measured
using Kron-like elliptical aperture magnitudes (\cite{Kron80}), with a
minimum Kron radius of 1.2~arcsec. They are corrected for
the galactic extinction estimated at the center of the VVDS-0226-04
field. For a large fraction of the field we have also U band data,
taken at the ESO 2.2m telescope and reaching a limiting magnitude of
$U_{AB} \sim 25.4$ (\cite{Radovich04}).

For the VVDS-CDFS, we have used the EIS $I$-band
photometry and astrometry (\cite{Arnouts01}) for our target selection,
and the multi-color $U$, $B$, $V$, $R$, and $I$ photometric catalogue
from the COMBO-17 survey (\cite{Wolf04}). 

The VVDS redshift survey uses the high multiplex capabilities of the
 VIMOS instrument installed at the Nasmyth platform of Melipal of the
 VLT-ESO in Paranal (Chile). The spectroscopic observations were
 obtained during two runs between October and December 2002. The
 spectroscopic targets were selected from the photometric
 catalogues using the VLT-VIMOS Mask Preparation Software (VVMPS;
 \cite{Bottini05}2005). The spectroscopic multi-object exposures 
 were reduced using the VIPGI tool (\cite{Scodeggio05}). The sample of
 spectroscopic redshifts obtained in the VVDS-CDFS is described in
 \cite{LeFevre04_II}(2004b) and the sample obtained in the VVDS-F02 is
 described in \cite{LeFevre05}(2005). A total of 11,034 spectra
 were acquired as primary targets in the two fields. The range of
 magnitude of the observed objects is $17.5 \le I_{AB} \le 24.0$. The
 deep spectroscopic sample (VVDS-0226-04+VVDS-CDFS) consists of
 6582+1258 galaxies, 623+128 stars and 62+9 QSOs with secure
 spectroscopic identification, i.e. quality flags 2, 3, 4 and 9 (flags
 2, 3, 4 correspond to redshifts measured with a confidence level of
 75\%, 95\%, 100\%, respectively; flag 9 indicates spectra with a
 single emission line). 1439+141 objects have an uncertain redshift
 measurement, i.e. quality flag 1 (flag 1 corresponds to a confidence
 level of 50\% in the measured redshift). 690+102 objects have no
 spectroscopic identification, i.e. quality flag 0.  This sample
 covers $\sim 1750+450$~arcmin$^2$, with a median redshift of about
 0.76. The 1$\sigma$ accuracy of the redshift measurements is
 estimated at $\sim$0.001 from repeated VVDS observations
 (\cite{LeFevre05}2005).

\section{Treatment of unidentified sources}

In the estimate of the luminosity function, we introduce a statistical
weight $w_i$, associated with each galaxy $i$ with a secure redshift
measurement. This weight corrects for the non-observed
sources and those for which the spectroscopic measurement failed
(unidentified sources).  This method yields the best statistical
estimate of the total number of galaxies with the same properties as
galaxy $i$, in the full field of view sampled by the spectroscopic
data. The statistical weight $w_i$ is the product of :
\begin{itemize}
\item the weight $w_i^{TSR}$, inverse of the {\it Target Sampling
    Rate} (hereafter {\it TSR}). The {\it TSR} is the fraction of
  objects in the photometric catalogue that have been
  spectroscopically targeted.  It can be a constant or a function of a
  number of parameters according to the strategy adopted in selecting
  the spectroscopic targets.
\item the weight $w_i^{SSR}$, inverse of the {\it Spectroscopic
    Success Rate} (hereafter {\it SSR}). The {\it SSR} is the fraction
  of the spectroscopically targeted objects for which a secure
  spectroscopic identification has been obtained. In this paper, the
  LFs are computed using galaxies with secure redshift measurements,
  i.e. with spectroscopic quality flag 2, 3, 4 and 9.  The {\it SSR}
  is the ratio between the number of objects with high quality flag 2,
  3, 4 and 9 and the total number of spectra (quality flags 0, 1, 2, 3, 4
  and 9). The correct estimate of the {\it SSR} is not trivial,
  because it can be a function of a large number of parameters, like,
  for example, magnitude, surface brightness, redshift and spectral
  type of the objects.
\end{itemize}

\subsection{The {\it TSR} and its associated weight $w^{TSR}$}

\begin{figure}[ht]
\centering \includegraphics[width=9cm]{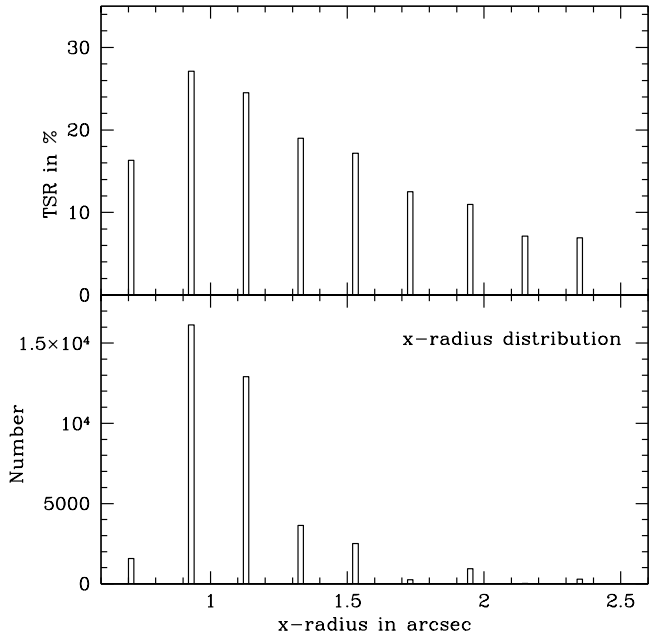}
\caption{ Top panel: {\it Target Sampling Rate} as a function of the
  x-radius for the VVDS-0226-04 field. Bottom panel: x-radius
  distribution in the VVDS-0226-04 photometric catalogue at $17.50 \le
  I_{AB} \le 24.0$.}
\label{figureTSR}
\end{figure}

The VVDS strategy in selecting spectroscopic targets has been to
select targets quasi-randomly from the photometric catalog, thus
minimizing any bias in sampling the galaxy population. In the random
selection process, the VVMPS tool (\cite{Bottini05}2005) uses the
information about the size of the objects in order to maximize the
number of slits per VIMOS pointing. As a consequence, the final
spectroscopic sample presents a bias with respect to the photometric
one, with large objects being under represented (see
\cite{Bottini05}2005, for a discussion). The parameter used by VVMPS
to maximize the number of slits is the x-radius, which is the
projection of the angular size of each object on the x-axis of the
image, corresponding to the direction in which the slits are
placed. The x-radius is defined as x-radius=$(n+0.5)\times 0.204$,
where 0.204 is the pixel size of the image expressed in $arcsec$ and
$n$ is an integer corresponding to the size of the object in
pixels. The {\it TSR} in the VVDS-0226-04 is shown as a function of
the x-radius in the top panel of Fig.\ref{figureTSR}. The {\it TSR}
runs from $\sim 25$\% for the smallest objects, to $\sim 10$\% for the
largest ones. As shown in the bottom panel of Fig.\ref{figureTSR}, a
large fraction of the total population ($\sim 75$\%) is targeted with
$TSR \sim 25$\%. The under-sampling of the largest objects (x-radius
$>$ 1.7) concerns less than $4$\% of the total population, targeted
with $TSR < 15$\%. Since the x-radius is the only parameter used to
maximize the number of slits, the correction to be applied in order to
correct for this bias is well defined and corresponds to using the
weight $w_i^{TSR}=1/TSR(r_i)$, where $r_i$ is the x-radius of the
galaxy $i$.

\subsection{The {\it SSR} and its associated weight $w^{SSR}$}

\begin{figure}
\centering \includegraphics[width=9cm]{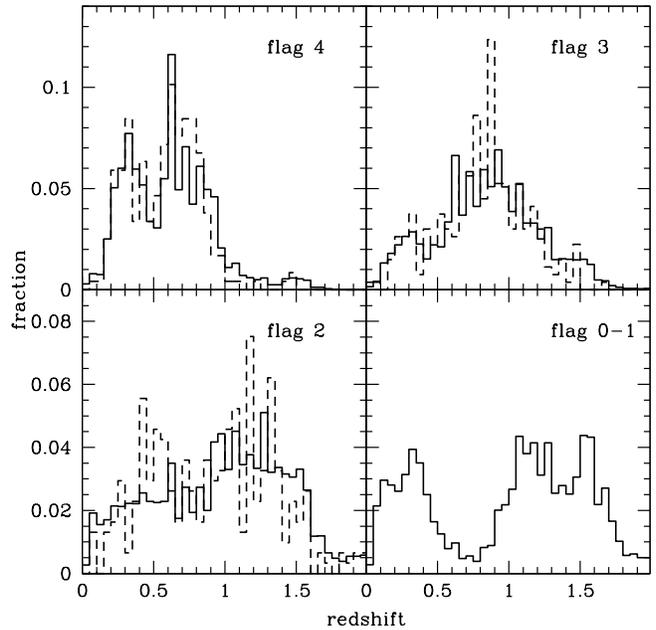}
\caption{Redshift distributions normalized to unity for
  each spectroscopic quality flag. For high quality flags 2, 3, 4, we
  show both the spectroscopic redshift distributions (dashed lines)
  and the photometric redshift distributions (solid lines) on the same
  area. For quality flags 0 and 1, we show only the redshift distribution
  estimated using the photometric redshifts.}
\label{figureDistz}
\end{figure}

\begin{figure}
\centering 
\includegraphics[width=9cm]{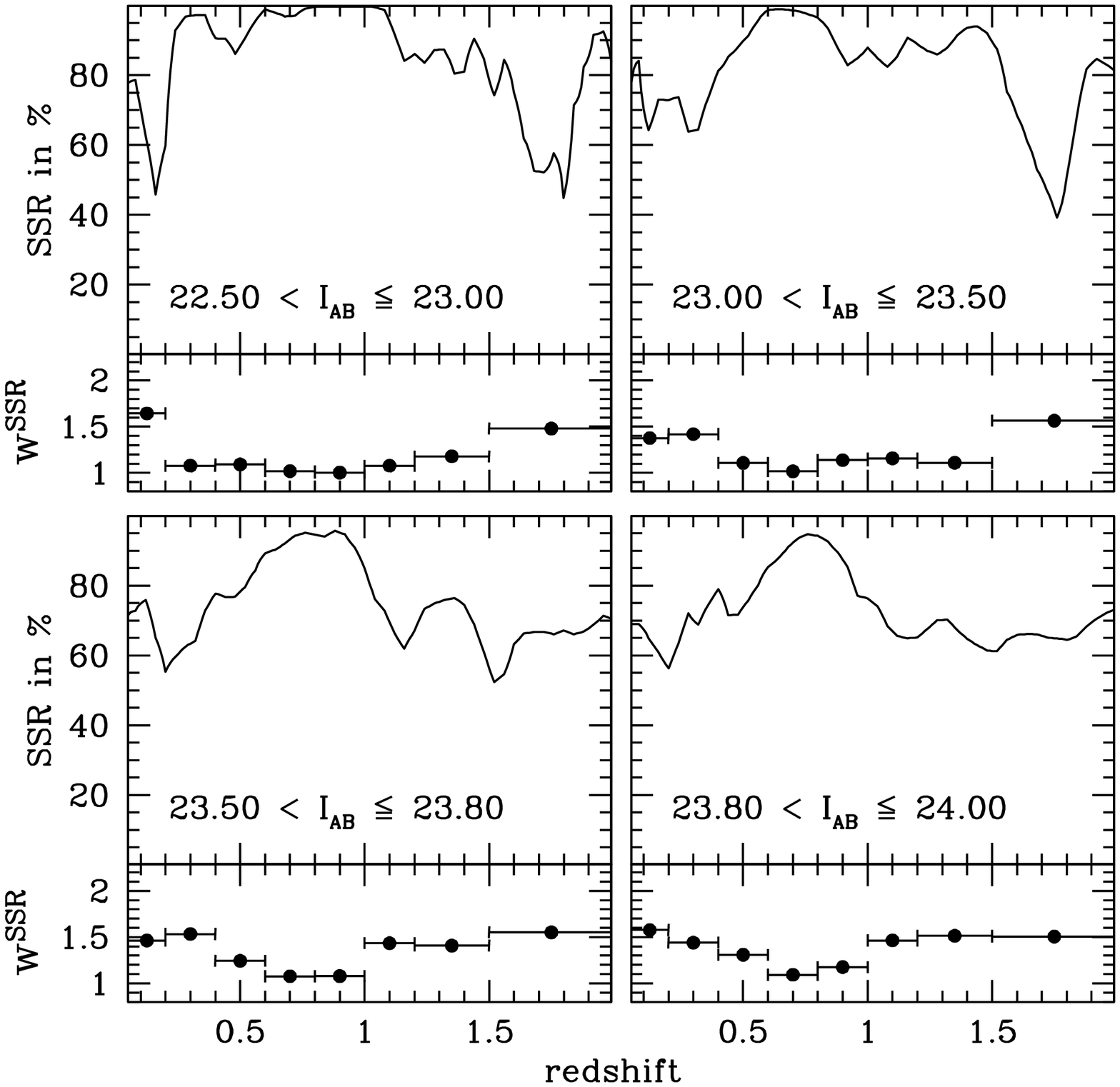}
\caption{ {\it Spectroscopic Success Rate} as a function of redshift and per 
apparent magnitude bin. The associated weight is shown in the bottom
of each panel.}
\label{figureSSR}
\end{figure}

The second weight to be used in the estimate of the LF is $w_i^{SSR}$,
which is the inverse of the {\it SSR}. In Fig.16 of
\cite{LeFevre05}(2005) it is shown that the {\it SSR} is, as
expected, a function of the $I_{AB}$ apparent magnitude.  The SSR is
greater than 90\% for $I_{AB} < 22.0$ and smoothly decreases down to
$\sim 70$\% in the faintest half a magnitude bin. In a first
approximation, we could use this {\it SSR} distribution to derive
$w_i^{SSR}$ as a function of the $I_{AB}$ apparent magnitude. However,
this procedure implies that the objects with quality flags 0 and 1
belong to the same population of the objects with a secure
spectroscopic identification (flags 2, 3 and 4). In particular, it
implies that they have the same redshift distribution. The redshift
distributions of galaxies with quality flags 4, 3 and 2 are shown in
Fig.\ref{figureDistz}. The distributions for each flag are clearly
different, reflecting the fact that the quality flag is related not
only to the signal-to-noise of the spectrum, but also to the number
and the strength of identifiable spectral features. This suggests that
the galaxies with quality flags 0 and 1 are likely to have a different
redshift distribution. If that is the case, we would not
be allowed to use $w_i^{SSR}$ as a function of magnitude only.

Therefore, making use of the multi-color properties of our sample, we
have analyzed the distribution of photometric redshifts for the
spectroscopic targets with flag 0 and 1. For this analysis we have
used only a subset area of the VVDS-0226-04 field, with $\sim 1100$
spectra, in which, in addition to the $U$ photometry
(\cite{Radovich04}), we have also $J$ and $K$ photometry
(\cite{Iovino05}). We have restricted this analysis to the
area with near-infrared data, since near infrared photometry allows
 us to estimate robust photometric redshifts at
least up to $z\sim 2$ (see \cite{Bolzonella05} for a detailed
description of the method). A redshift probability distribution
function (hereafter {\it PDFz}) is estimated for each object of the
spectroscopic sample, using the photometric redshift code of {\it Le
Phare}\footnote{www.lam.oamp.fr/arnouts/LE\_PHARE.html} (Arnouts \&
Ilbert). We sum the normalized {\it PDFz} of all galaxies to estimate the
expected redshift distribution (\cite{Arnouts02}). The stars are
removed from the sample on the basis of their spectral identification
if they have high quality spectroscopic flags, or on the
basis of photometric criteria for the low quality flags
(\cite{Bolzonella05}). The estimated redshift distribution of galaxies
with quality flag 0 and 1 is shown in the bottom right panel of
Fig.\ref{figureDistz}. As expected, the estimated redshift
distribution of the low quality flag galaxies differs from the
redshift distribution of high quality flag galaxies, while the
distributions of the photometric and spectroscopic redshifts for
galaxies with flag $ \ge $ 2 are consistent with each other.

We have then derived the {\it SSR} in various bins of apparent
magnitude as the ratio between the estimated redshift distribution of
high quality flag galaxies (quality flags 2, 3, 4, 9) and the
estimated redshift distribution of all galaxies (quality flags 0, 1,
2, 3, 4, 9). This {\it SSR} is shown in Fig.\ref{figureSSR} as a
function of redshift in four apparent magnitude bins for $I_{AB} \ge
22.5$ (at brighter magnitudes, the {\it SSR} is close to unity).
Fig.\ref{figureSSR} clearly shows that the global {\it SSR} indeed
decreases for fainter apparent magnitude bins and it varies
significantly with redshift. The shape of the {\it SSR} is similar in
all magnitude bins showing a maximum efficiency in the redshift
measurement at $z\sim 0.7$ and a minimum {\it SSR} for $z < 0.5$ and
$z > 1.5$. The dependence of the {\it SSR} on the redshift is related
to the presence of the [O II] line, and/or the Balmer break within the
observed spectral window 5500\AA $\le \lambda \le$ 9500\AA. The weight
$w^{SSR}$ is shown in the bottom of each panel in
Fig.\ref{figureSSR}.  The weight is binned in redshift in order to
limit the statistical noise.

At $z > 1.5$, the uncertainties on our weight are large due to the
smaller number of galaxies, and to the uncertainties on the
photometric redshifts at such redshifts (\cite{Bolzonella05}). We
could also perform an other estimate of the weight at such high
redshifts, using the spectroscopic redshifts of the spectra with
quality flag 1 (50\% of confidence level) and assuming that quality
flag 0 objects have the same redshift distribution. Applying this
method, we find {\it SSR}$\sim$20-30\% above $z > 1.5$, which provides
a weight $\sim$2-3 times greater than the weight estimated with the
photometric redshift method. The value of $w^{SSR}$ at $z>1.5$ will be
refined in future analysis (Paltani et al., in preparation), using simulations and
new spectroscopic VIMOS observations with a blue grism.

We apply the weights derived from this analysis to the whole
sample, making the assumption that the subset area, from which they
have been derived, is representative of our ability to measure a
redshift.  We assign to each galaxy, a weight $w_i^{SSR}$ that depends
both on the apparent magnitude $m_i$ and on the redshift $z_i$ of the
considered galaxy $i$.

To summarize, we have derived our statistical weights as the product of
$w_i=w_i^{TSR} \times w_i^{SSR}=1/TSR(r_i) \times 1/SSR(m_i, z_i)$.
This weighting scheme allows us to correct for:
\begin{itemize}
\item the {\it TSR}, taking into account the small under-sampling of the
  largest objects in the target selection;
\item the {\it SSR}, taking into account the dependence on the
  apparent magnitude and redshift.
\end{itemize}

\section{ALF applied to the VVDS}

We have measured the LF on the VVDS data, using our luminosity
function VVDS tool, named Algorithm for Luminosity Function (ALF).
The methods implemented in ALF are extensively described in the
Appendix. In this Section we briefly discuss two points which are
relevant for a better understanding of our treatment of the data
in this paper.

\subsection{Effect of the $w^{SSR}$ weight on the LF estimates}

\begin{figure*}[ht]
  \centering \includegraphics[width=15cm]{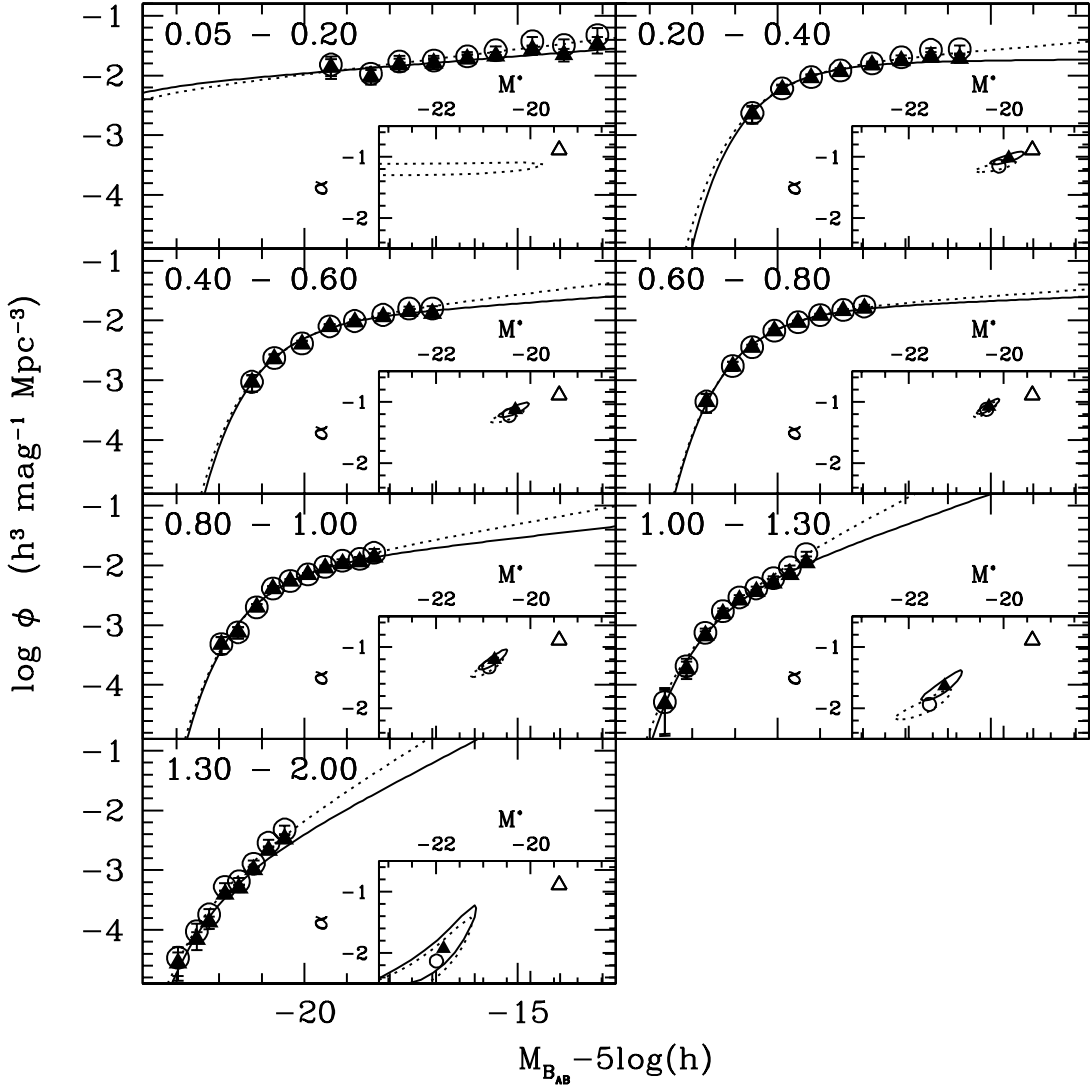}
\caption{Comparison between the `unweighted' LFs (solid triangles for
  1/V$_{\rm max}$, solid lines for STY) and our best LF
  estimates (`weighted') (open circles for 1/V$_{\rm max}$, dotted
  lines for STY) in the $B$ rest-frame band. The best STY fits and the
  associated 90\% error contours are shown as insets with the same
  symbols. We report in each inset the SDSS local estimate with open
  triangles.}
\label{figureWeight}
\end{figure*}

As shown in Section 3, while the weight $w^{TSR}$ is fully understood and
well established, the derivation of the weight $w^{SSR}$ is less
direct and subject to more uncertainties. To quantify the effect of
$w^{SSR}$ on our LF estimate, we have also derived  an `unweighted' LF,
in which no correction for the {\it SSR} is applied (i.e. $w^{SSR}=1$).
The `weighted' and `unweighted' LFs are shown in Fig.\ref{figureWeight},
in the $B$ rest-frame band.

Since the galaxies with flag 0 and 1 can not be ignored in the LF
estimate, the `unweighted' LF is by definition a lower limit of the
LF. However, given the relatively small fraction of galaxies with flag
0 and 1, the difference in the overall normalization of the two LFs is
small. From Fig.\ref{figureWeight}, we see that the main effect of
$w^{SSR}$ is to steepen the slope of the `unweighted' LF. In all the
redshift bins, the parameter $\alpha$ of the weighted LF is smaller
(i.e. steeper slope), by less than 0.2 up to $z=1.0$, by less than 0.3
in the two higher redshift bins. This effect is clearly expected,
since the galaxies with flag 0 and 1, which are included in the LF
estimate through the weight, are mainly faint galaxies close to our
magnitude limit.  Since $\alpha$ and $M^*$ are correlated, the
steepening of the slope with the weight produces also a brightening of
$M^*$, less than 0.25 up to $z=2$.

\subsection{Estimate of the LF from a homogeneous galaxy population}

Ilbert et al.(2004) have shown that the estimate of the global LF can
be biased, mainly at its faint-end, when the band in which the global
LF is measured is far from the rest-frame band in which galaxies are
selected. This is because different galaxy types have
different absolute magnitude limits, because of different
k-corrections. In each redshift range, we avoid this bias
in our estimates of the LF by using only galaxies within the absolute
magnitude range where all the SEDs are potentially observable. We
perform only the 1/V$_{\rm max}$ estimate on the whole absolute
magnitude range. This estimator leads to an under-estimate of the LF
in the absolute magnitude range fainter than this `bias'
limit (\cite{Ilbert04}2004), providing a lower limit of the LF
faint-end.

\section{Results}

\begin{figure*}[ht]
\centering 
\includegraphics[width=15cm]{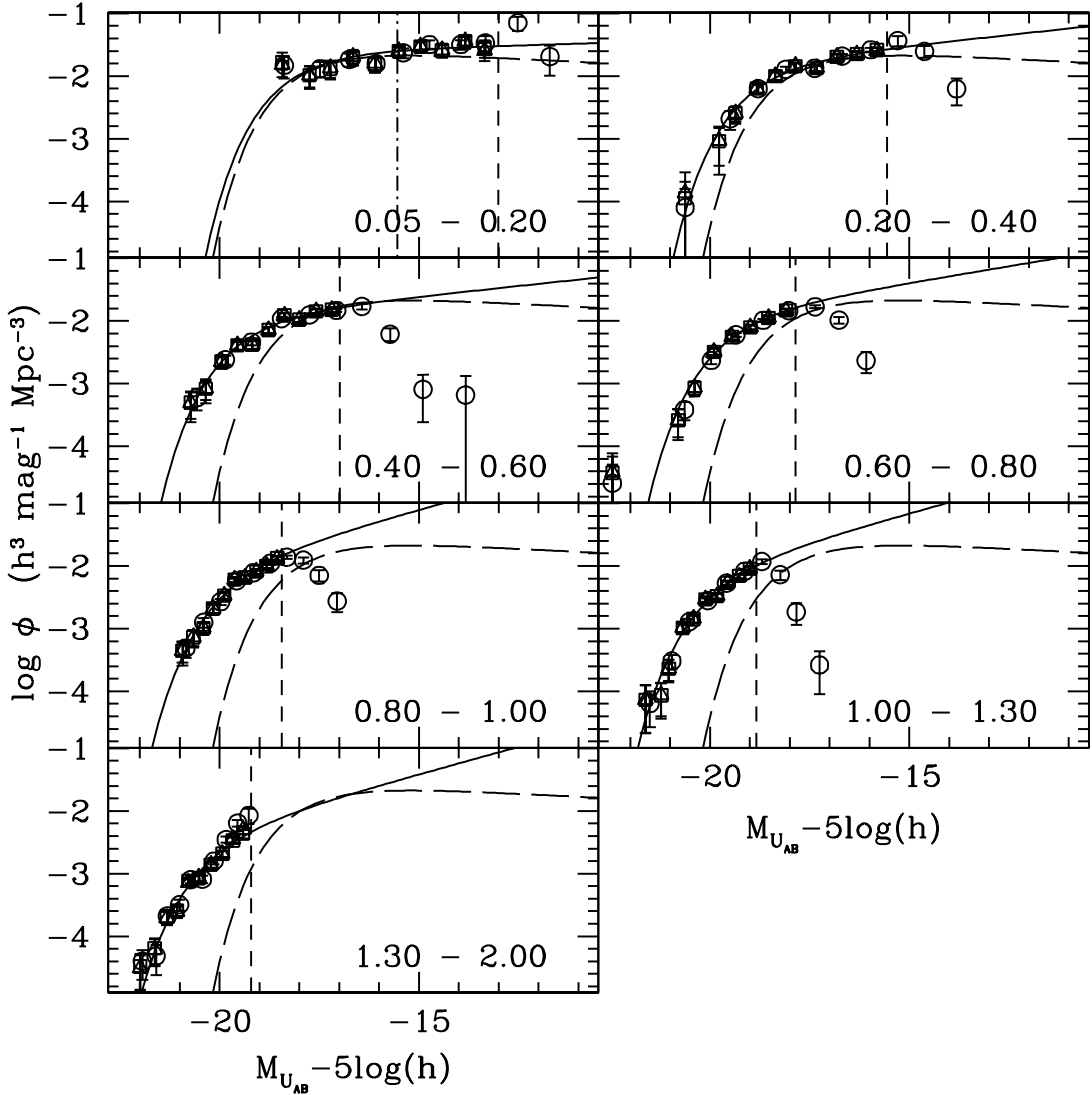}
\caption{Estimate of the global LF in the $U$ band from $z = 0.05$ to 
  $z = 2$. The estimate is derived using the weighting scheme
  described in Section 3. We adopt the following symbols for the
  various estimators: circles for the 1/V$_{\rm max}$, triangles for
  the SWML, squares for the C$^+$ and solid lines for the STY. The
  vertical short-dashed lines show the faint absolute magnitude limits
  considered in the STY estimate (see Section 4.2). In each panel, we
  show also the local LF derived by the SDSS (long dashed lines). The
  vertical dot-dashed line in the redshift bin 0.05-0.2, corresponds
  to the faint absolute magnitude limit surveyed by the SDSS.}
\label{figureLFU}
\end{figure*}

\begin{figure*}
\begin{tabular}{ll}
\includegraphics[width=8.8cm]{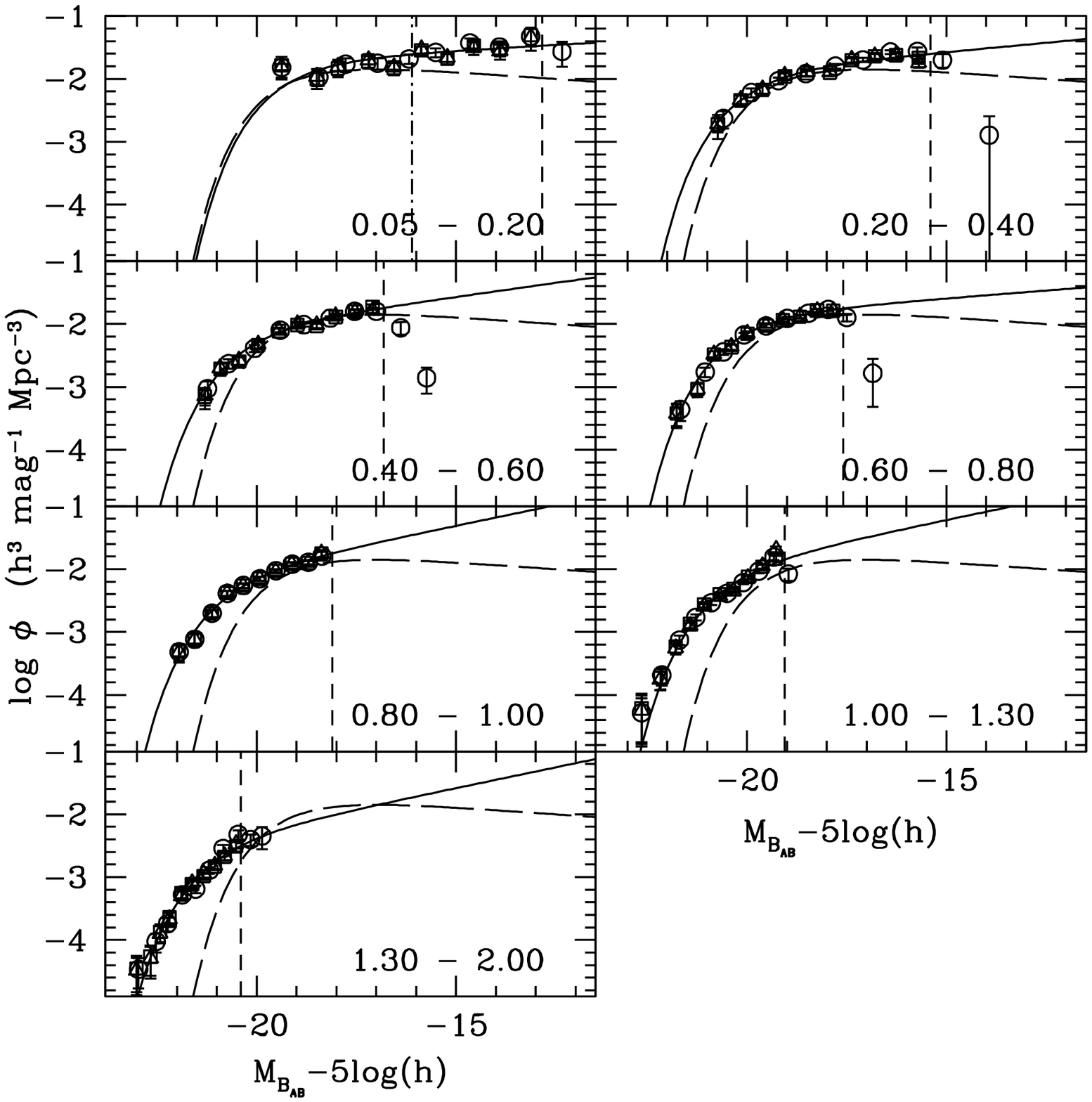} &
\includegraphics[width=8.8cm]{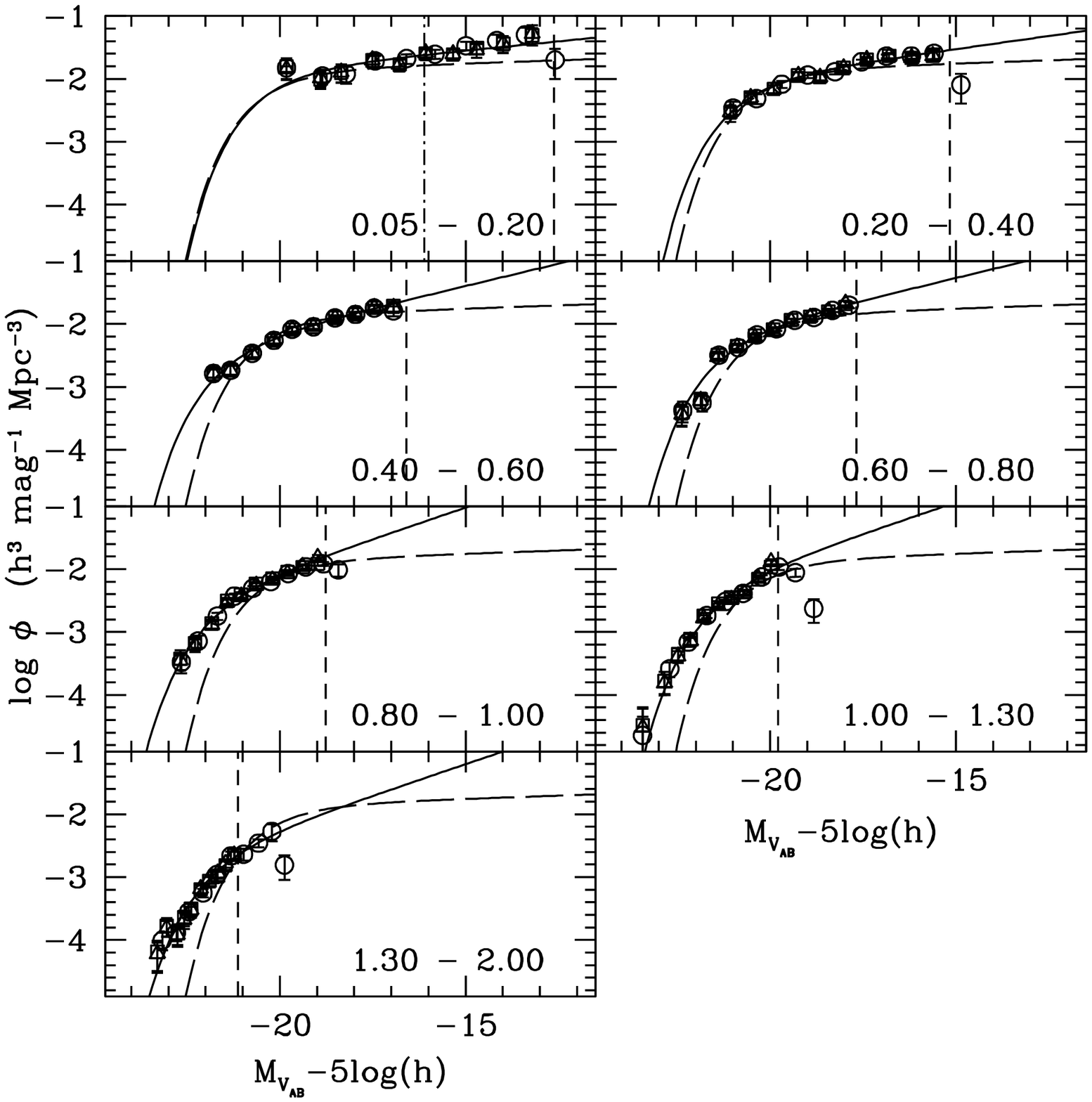} \\
\includegraphics[width=8.8cm]{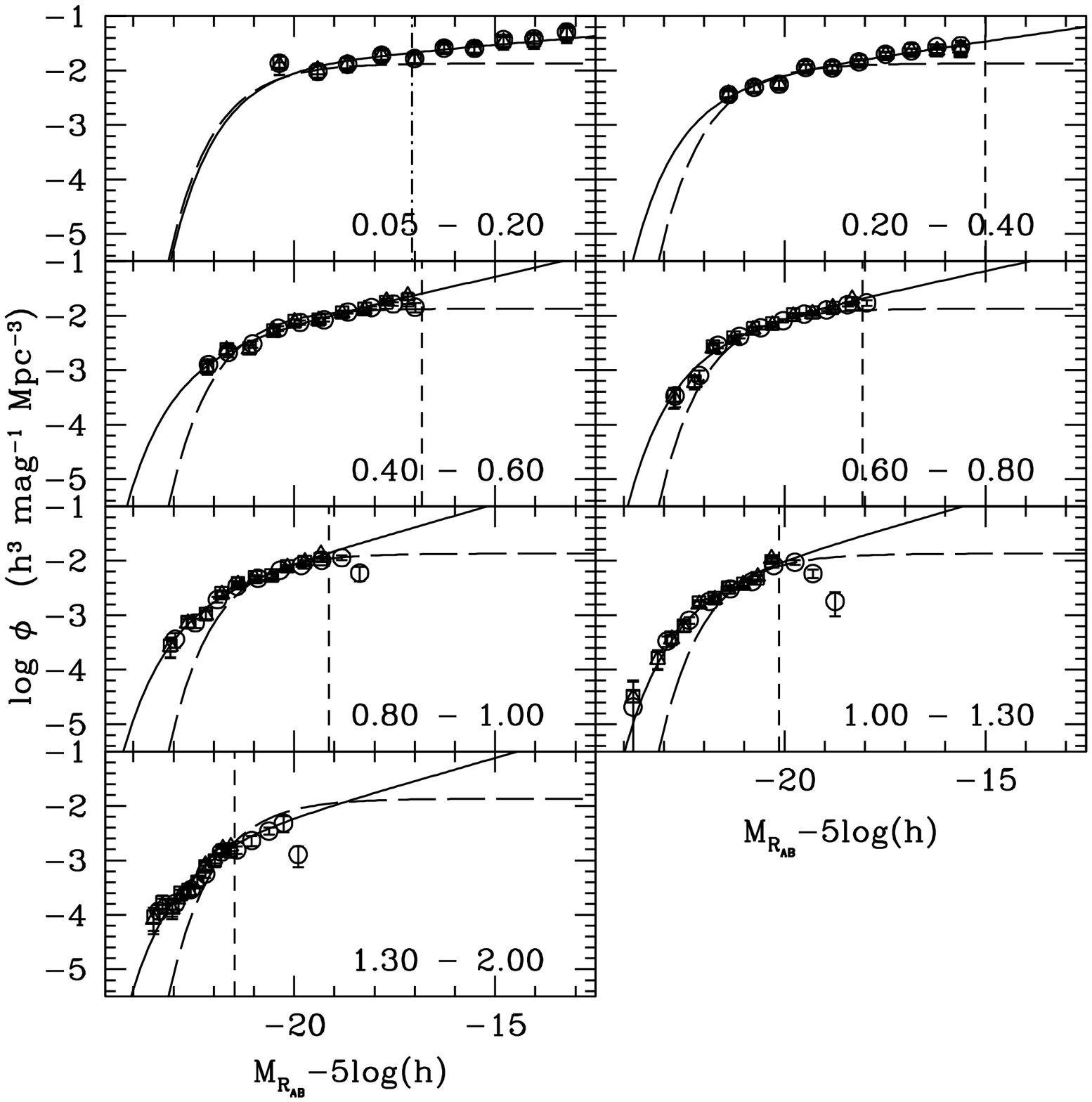} &
\includegraphics[width=8.8cm]{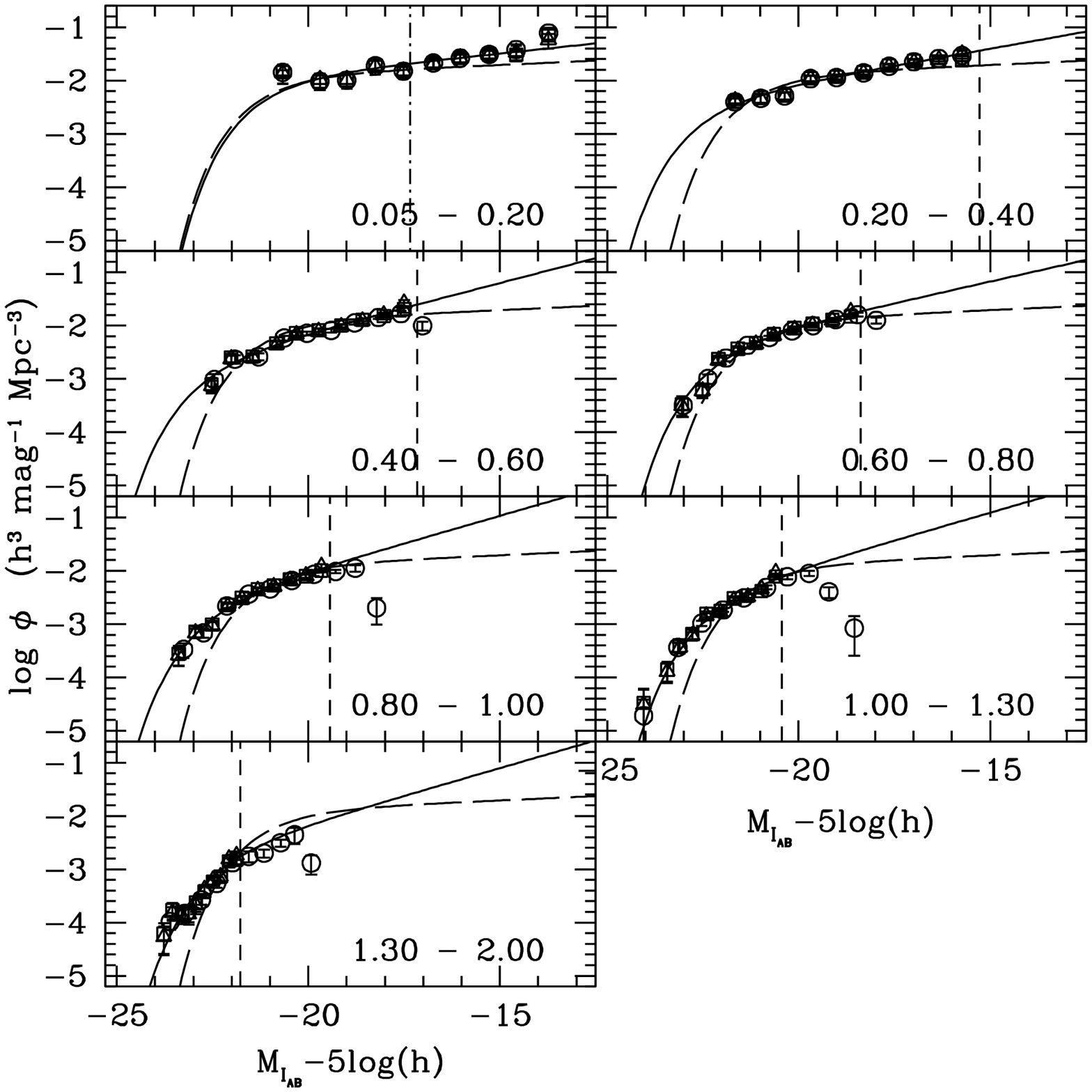}
\end{tabular}
\caption{Same symbols as in Fig.\ref{figureLFU}, in the rest-frame 
  band $B$ (upper-left), $V$ (upper-right), $R$ (lower-left) and
  $I$ (lower-right).}
\label{figureOtherLF}
\end{figure*}

\begin{figure*}[ht]
  \centering \includegraphics[width=15cm]{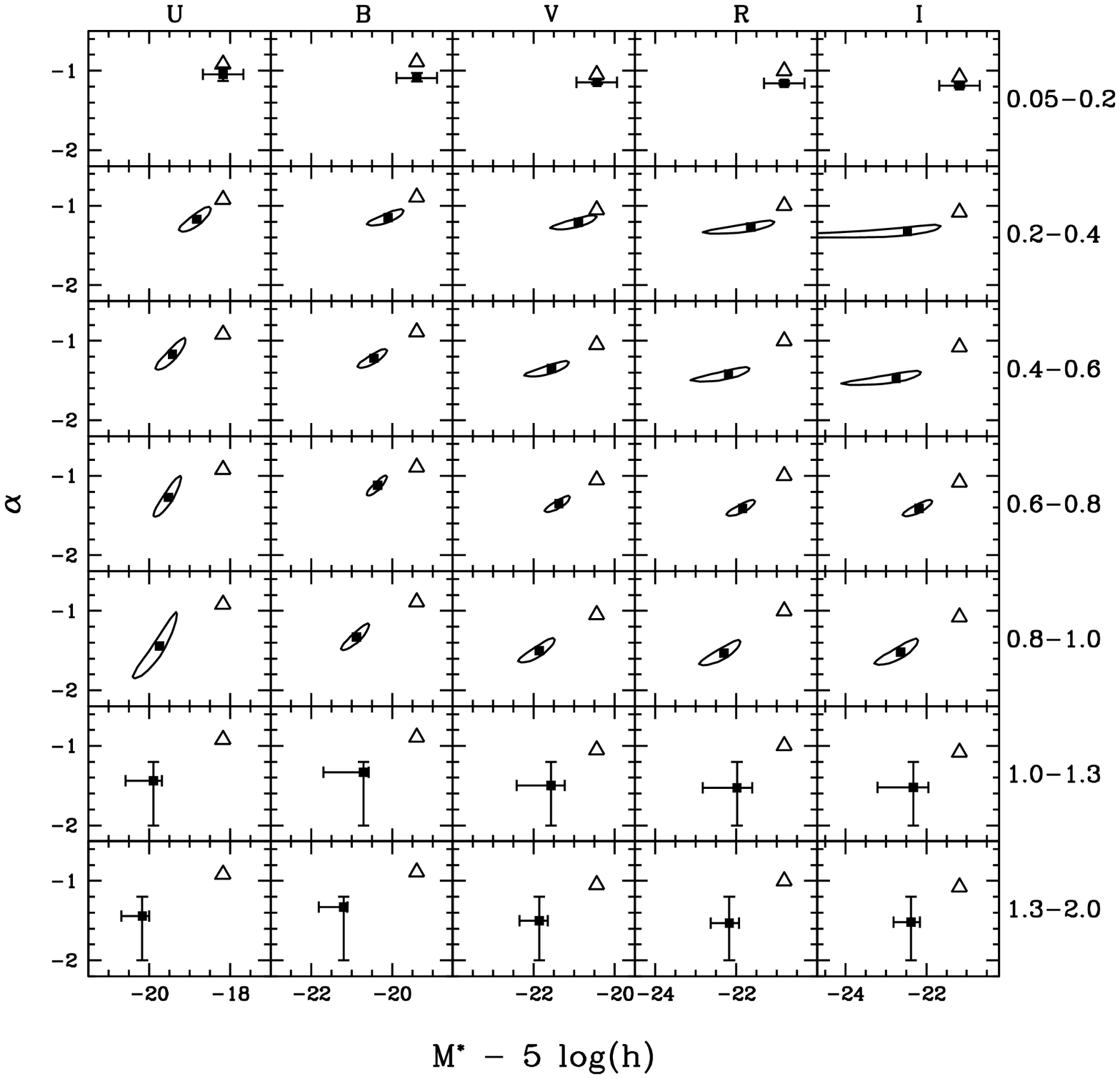}
\caption{ $M^*-\alpha$ error contours, at 90\% confidence level
  ($2\Delta ln\mathcal L=4.61$), obtained with the STY estimator. The
  solid squares are the STY best estimates. Error contours are not
  drawn when one of the two values has been fixed (see
  text). When $\alpha$ is fixed, we estimate the range of
  allowed $M^*$ values, varying $\alpha$ between $\alpha=-2.0$ and
  $\alpha=-1.2$. When $M^*$ is fixed, we vary $M^*$ by 0.5
  around the set value. The error contours are shown for the $U$, $B$,
  $V$, $R$ and $I$ band LFs from the left to the right panels,
  respectively. The panels from the top to the bottom correspond to
  the eight redshift bins, from $z=0.05$ to $z=2$.  We report with
  open triangles the SDSS local values (the error bars are included
  inside the symbols).}
\label{figureEll}
\end{figure*}

The global LFs are computed up to $z = 2$ in the five standard bands
$U$, $B$, $V$, $R$, $I$ ($U$ Bessel, $B$ and $V$ Johnson, $R$ and $I$
Cousins). The LFs are computed using the weighting scheme described in
Section 3. The $U$-, $B$-, $V$-, $R$- and $I$-band LFs are displayed
in Fig.\ref{figureLFU} and in Fig.\ref{figureOtherLF}.
In Fig.\ref{figureEll}, we plot the STY best estimate of $\alpha-M^*$,
and the associated error contours. For each band and each redshift
bin, the Schechter parameters and the corresponding one sigma errors
measured with the STY estimator are listed in
Table~\ref{table1}.

\subsection{Local LF at $z = 0.1$}

The local LF derived with the VVDS sample refers to the redshift bin
[$0.05-0.2$]. The average redshift in this bin ($<z> \sim 0.14$) is
directly comparable to the average redshift of galaxies in local
surveys with a brighter limiting magnitude, like the SDSS ($<z> \sim
0.1$).  Due to the bright apparent magnitude cutoff of the VVDS sample
($I_{AB} \ge 17.5$), the $M^*$ parameter of the STY fit in this
redshift bin is essentially unconstrained. Therefore, we set the $M^*$
parameter to the local value derived by \cite{Blanton03}(2003). The
LFs of the SDSS are expressed in the bands $^{0.1}u$, $^{0.1}g$,
$^{0.1}r$, $^{0.1}i$, $^{0.1}z$ (\cite{Fukugita96}) blue-shifted at $z
= 0.1$, which correspond roughly to the $U$, $B$, $V$, $R$, $I$ bands
of our standard system. In order to check if the absolute
magnitudes estimated in the SDSS band system and in our standard band
system are comparable, we have estimated the absolute magnitudes in
the filters $^{0.1}u$, $^{0.1}g$, $^{0.1}r$, $^{0.1}i$ and $^{0.1}z$
from the apparent magnitudes measured in the instrumental system
(using the formulae A.1 and A.2 given in appendix A). The average
difference between the absolute magnitudes computed in $B$, $V$, $R$,
$I$ and in $^{0.1}b$, $^{0.1}v$, $^{0.1}r$, $^{0.1}i$ bands are less
than 0.05. The difference is more significant in the $U$ band ($\Delta
M \sim 0.25$). We have therefore converted $M^*_{SDSS}(^{0.1}u)$ to
our band with the relation
$M^*_{SDSS}(U)=M^*_{SDSS}(^{0.1}u)-0.25$. The local values of $\alpha$
and $\phi^*$, with $M^*$ set to the SDSS value, are listed in 
Table~\ref{table1} and the LFs are shown in Fig.\ref{figureLFU} and
Fig.\ref{figureOtherLF}. Even if the volume surveyed by the VVDS in
the first redshift bin is approximately one thousand times smaller
than the volume surveyed by the SDSS, the estimates of the
local LFs produced by the VVDS and the SDSS are in good agreement in
the magnitude range in common to both surveys. However, in all
the bands, the VVDS best fit slope is steeper than the
SDSS slope.  The larger difference is in the $B$ band, where it is
formally significant at $\sim 3\sigma$ level 
($\alpha_{SDSS} = -0.89\pm0.03$ while $\alpha_{VVDS} =
-1.09\pm0.05$). In this band the VVDS slope is instead consistent
with that derived by \cite{Norberg02}(2002) from the 2dFGRS.  Even if
the number of objects in the VVDS is smaller than in the SDSS, the
faint-end slope of the LF is better constrained by the VVDS because it
samples the local galaxy population about 3-4 magnitudes deeper than
the SDSS.  The steeper slope observed in the VVDS cannot be due to the
effect of the applied weights since also the `unweighted' LF, which
under-estimates the slope (see Section 4.1), has a steeper best fit
slope than the SDSS ($\alpha_{VVDS} = -1.02\pm0.05$ in the
$B$-band). The inclusion of a fit for simple luminosity and number
evolution in the LF estimate, using the maximum likelihood estimator
proposed by \cite{Blanton03}(2003), could also produce a flatter
slope.

\subsection{LF evolution up to $z = 2$}

\begin{figure}[h]
\centering
\includegraphics[width=9cm]{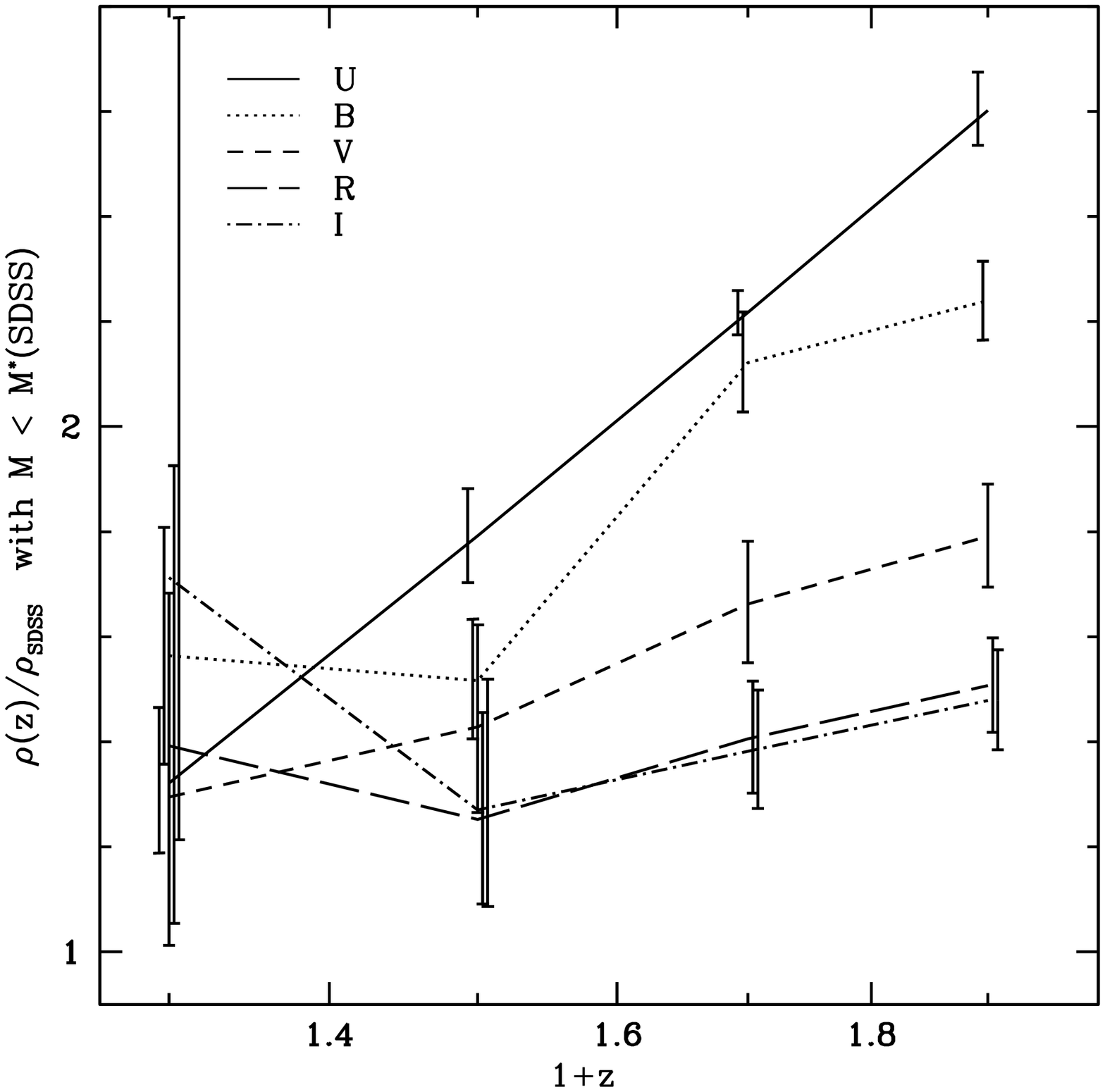}
\caption{Evolution with redshift of the ratio between the density
of galaxies brighter than $M^*_{SDSS}(z=0.1)$ and the SDSS local value.}
\label{figureDens}
\end{figure}

The VVDS allows us to quantify the galaxy evolution within
a single sample and with the same selection function, over a wide
redshift range. From $z = 0.05$ up to $z = 2$, the evolution
of the bright part of the LFs is clearly evident from all
non-parametric estimators shown in Fig.\ref{figureLFU} and
Fig.\ref{figureOtherLF}. It also appears to be a function of the
considered rest-frame wavelength. This can be quantified using the
Schechter parameters measured with the STY estimator, as done below.

To quantify the strength of the evolution with redshift, we have
 derived the density of galaxies brighter than the corresponding local
 value of $M^*$: $$ \rho(M<M^*_{SDSS})=\int_{-\infty}^{M^*_{SDSS}}
 \phi(M) dM,
$$
where $M^*_{SDSS}$ is the local value estimated by
\cite{Blanton03}(2003). This measurement quantifies the global
evolution of the bright part of the LFs, in shape and in
normalization. In all bands and up to $z \le 1$, $M^*_{SDSS}$ is
brighter than the faintest limits used for the STY estimate. We have
therefore limited this analysis to z $\le $ 1, in order to avoid
extrapolation of the LF beyond the last measured point. In
Fig.\ref{figureDens} we show the evolution with redshift of the ratio
$\rho(M<M^*_{SDSS})/\rho_{SDSS}(M<M^*_{SDSS})$ for the five bands. As
the figure clearly shows, the density evolution of bright galaxies is
significantly dependent on the rest-frame wavelength, being stronger
at shorter wavelengths. In the $U$ band, $\rho(M<M^*_{SDSS})$
increases continuously and becomes $\sim 2.6$ times
greater than locally at $z = 1.0$, while in the $I$ band this factor
is only $\sim 1.5$ at $z = 1$.

\begin{figure}[h]
  \centering \includegraphics[width=19cm]{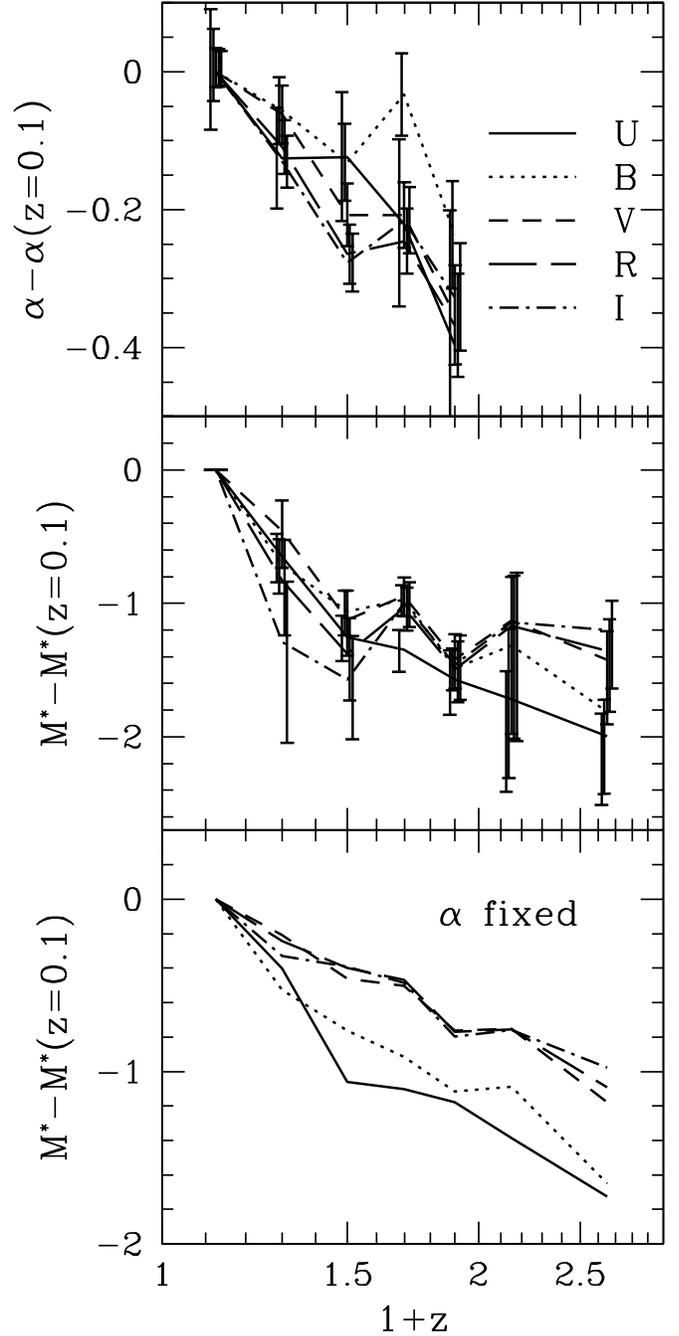}
\caption{ Evolution in the five bands of the parameter 
  $\alpha$ in the top panel and of the parameter $M^*$ in
  the second panel. The bottom panel shows the evolution of 
  $M^*$ with the slope fixed to the VVDS local value.}
\label{figureEvol}
\end{figure}

The evolution of the best fit $M^*$ as a function of redshift for the
five bands is shown in the central panel of Fig.\ref{figureEvol}. We
find that the characteristic magnitude M$^{*}$ of the whole population
strongly evolves. Up to $z=1$, the slope can still be
constrained reasonably well and we measure a brightening
of $1.57 \pm 0.26$, $1.48 \pm 0.17$, $1.41 \pm 0.22$, $1.49 \pm 0.25$
and $1.45 \pm 0.26$ magnitudes in the $U$, $B$, $V$, $R$ and $I$
rest-frame bands, respectively. Above $z=1$, the slopes are
set to the $\alpha$ value obtained in the redshift bin $0.8 \le z \le
1$ and we measure a brightening of about $2.0$, $1.8$, $1.4$, $1.3$,
$1.2$ magnitudes up to $z=2$. When $\alpha$ is fixed, we
estimate the range of allowed $M^*$ values, varying $\alpha$ between
two extreme values of the slope, $\alpha=-2.0$ and $\alpha=-1.2$. We
find a brightening included in the range $1.8-2.5$, $1.7-2.4$,
$1.2-1.9$, $1.1-1.8$ and $1.0-1.6$ magnitudes in the $U$, $B$, $V$,
$R$ and $I$ rest-frame bands, respectively. Also in this
representation, the evolution is stronger in the bluer rest-frame
bands.  Since $M^*$ and $\alpha$ are correlated and we have some
evidence that also $\alpha$ is changing with redshift (see below and
upper panel of Fig.\ref{figureEvol}), we have verified that the
observed evolution in $M^*$ is not induced by the change with redshift
of the $\alpha$ value. The bottom panel of
Fig.\ref{figureEvol} shows the best fit $M^*$ parameters 
derived by setting the value of $\alpha$ to the VVDS local value over
the entire redshift range.  Also in this case, a significant and
differential evolution of $M^*$ is seen, with $\Delta M^*\sim -1.7,
-1.6, -1.2, -1.1, -1.0$ up to $z\sim2$, in the
$U$, $B$, $V$, $R$, $I$ bands respectively. The
measurement of this brightening is slightly sensitive to the adopted
weighting approach and it is also measured, at a similar level, with
the `unweighted' LFs (see Fig.\ref{figureWeight}).

The upper panel of Fig.\ref{figureEvol} shows the best fit values of
$\alpha$ as a function of redshift. The one sigma error bars on
$\alpha$ take into account the correlation between $\alpha$ and $M^*$.
The data suggest a steepening of the slope with increasing
redshift. The measured variation of $\alpha$ between $z =0.05$ and $z
= 1$ is $\Delta\alpha \sim -0.3$, similar in all the bands.

\begin{figure}[h]
  \includegraphics[width=17cm]{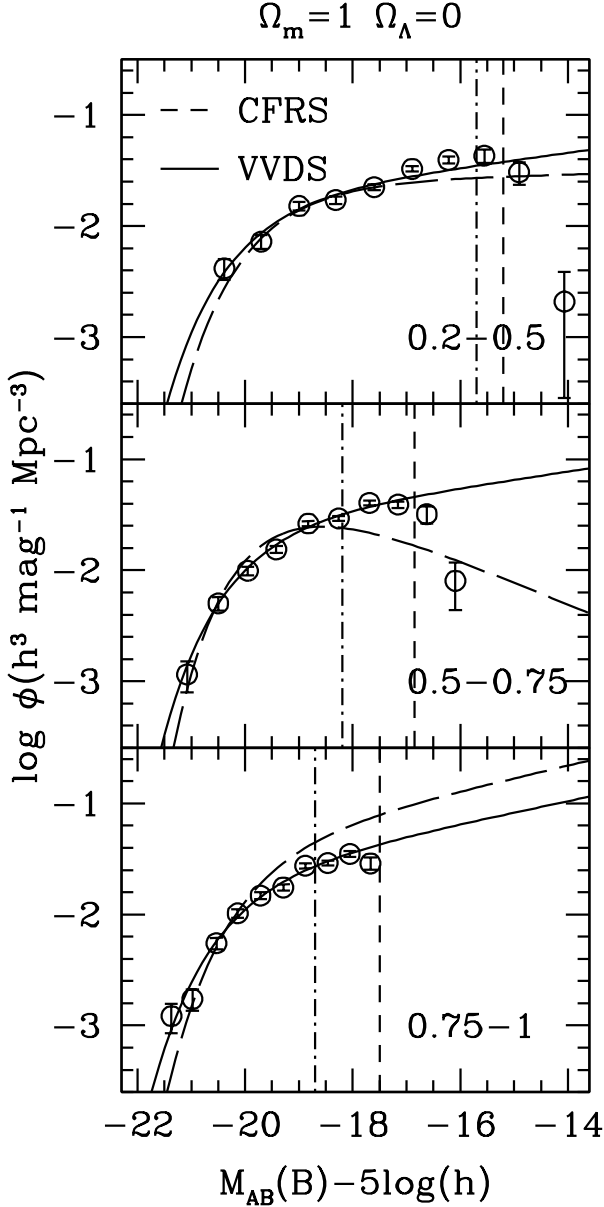}
\caption{Comparison between the CFRS and the
  VVDS global $B$-band LFs. The solid lines (STY) and the circles
  (1/V$_{\rm max}$) correspond to the VVDS estimates. The vertical
  short-dashed lines are the faint absolute magnitude limits
  considered in the STY estimates. The long-dashed lines correspond to
  the CFRS estimate. The vertical dot-dashed lines correspond to the
  faint absolute magnitude limits surveyed by the CFRS.}
\label{figureCFRS}
\end{figure}

\section{Comparison with previous redshift surveys}

\subsection{Comparison with the CFRS survey}

Lilly et al. (1995) have derived the global $B$-band LFs
of the Canada-France Redshift Survey (CFRS) up to $z \sim 1$. The CFRS
spectroscopic sample contains 591 redshifts of galaxies selected with
$17.5 \le I_{AB} \le 22.5$. The survey covers 125 arcmin$^2$ in five
separated fields. The VVDS deep spectroscopic sample is surveying the
galaxy population 1.5 magnitude fainter and the field of view is ten
times larger than the CFRS. The comparison between VVDS and CFRS
results in three redshift bins is displayed in Fig.\ref{figureCFRS}
(in the cosmology $\Omega_m=1$, $\Omega_\Lambda=0$, which was adopted
in the CFRS analysis). The estimated LFs for the two surveys are in
agreement up to the faintest absolute magnitude limits reached by the
CFRS. The slopes of the VVDS are, however, steeper than the CFRS
slopes (the difference is particularly significant in the redshift bin
]0.5-0.75]). The slopes estimated from the VVDS are clearly more
robust, since the VVDS is 1.5 magnitudes deeper and contains 10 times
more galaxies than the CFRS.

\begin{figure}[h]
  \includegraphics[width=17cm]{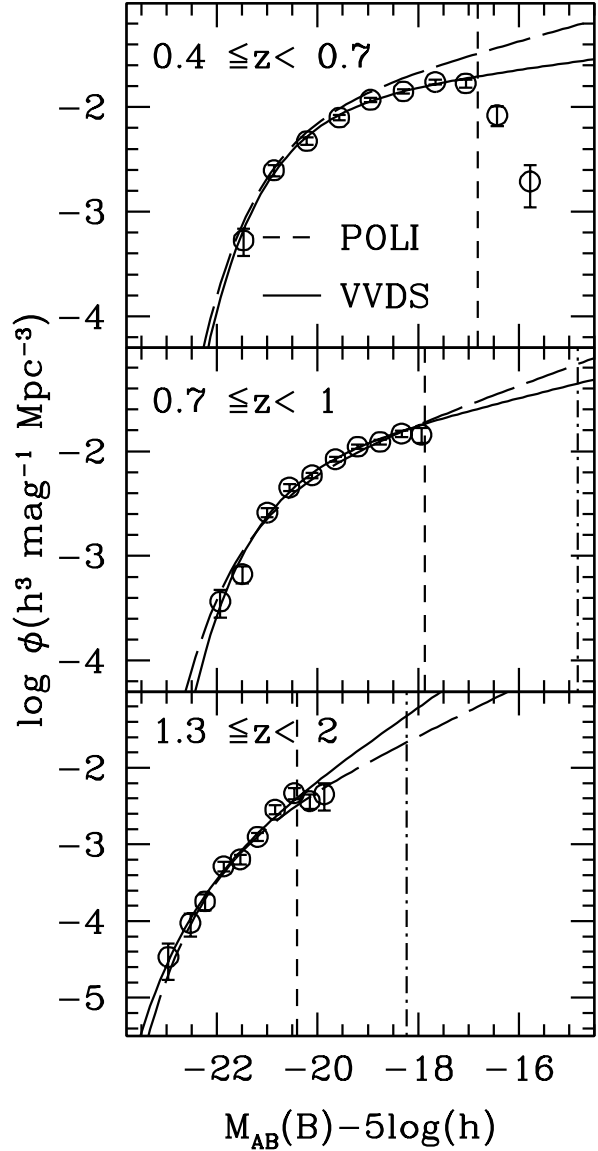}
\caption{Comparison between the global $B$-band LFs
  derived with the HDF data (\cite{Poli03}2003) and with
  the VVDS data.  The solid lines (STY) and the circles (1/V$_{\rm
  max}$) correspond to the VVDS estimates. The vertical short-dashed
  lines are the faint absolute magnitude limits considered in the STY
  estimates.  The long dashed lines correspond to the global LFs
  derived by \cite{Poli03}(2003). The vertical dot-dashed lines
  correspond to the faint absolute magnitude limits surveyed by the
  HDF data. }
\label{figurePoli}
\end{figure}

\subsection{Comparison with the HDF data}

Poli et al. (2003) have derived the global $B$-band LFs
from z = 0.4 up to z = 3.5, using a composite sample of 1541
I-selected galaxies down to $I_{AB} = 27.2$ and 138 K-selected
galaxies down to $K_{AB} = 25$. The faintest galaxies of this
composite sample are drawn from HDF North and South data. Data from
two additional fields (the CDFS and the field around the QSO
0055-269), on which the K20 spectroscopic survey is based
(\cite{Cimatti02}), have been added to constrain the bright-end of the
LF. Given the faintness of this sample, most of the
redshifts ($\sim 80\%$) are photometric redshifts. \cite{Poli03}(2003)
have derived the global $B$-band LF using the $I$-selected sample up
to $z = 1$ and the $K$-selected sample for $z \ge 1.3$. The
HDF data survey the LF faint-end about 2-3 magnitudes fainter than
the VVDS data. The LFs from \cite{Poli03}(2003) and the corresponding
VVDS LFs are shown in Fig.\ref{figurePoli} in three redshift bins. At
$z > 1.3$, \cite{Poli03}(2003) have derived the LF in the
redshift bin [1.3-2.5], that we compare here with our measurement in
the redshift bin [1.3-2]. As shown in
Fig.\ref{figurePoli}, there is an excellent agreement in the bright
part of the LF between the VVDS and the \cite{Poli03}(2003)
measurements, up to $z = 2$. In the faint part of the LF, the slope
estimated by \cite{Poli03}(2003) is slightly steeper ($\Delta \alpha
\sim 0.15$) than the slope estimated with the VVDS data in the
redshift bin [0.4-0.7].

\subsection{Comparison with the COMBO-17 survey}

\begin{figure*}[ht]
\begin{center}
\includegraphics[width=15cm]{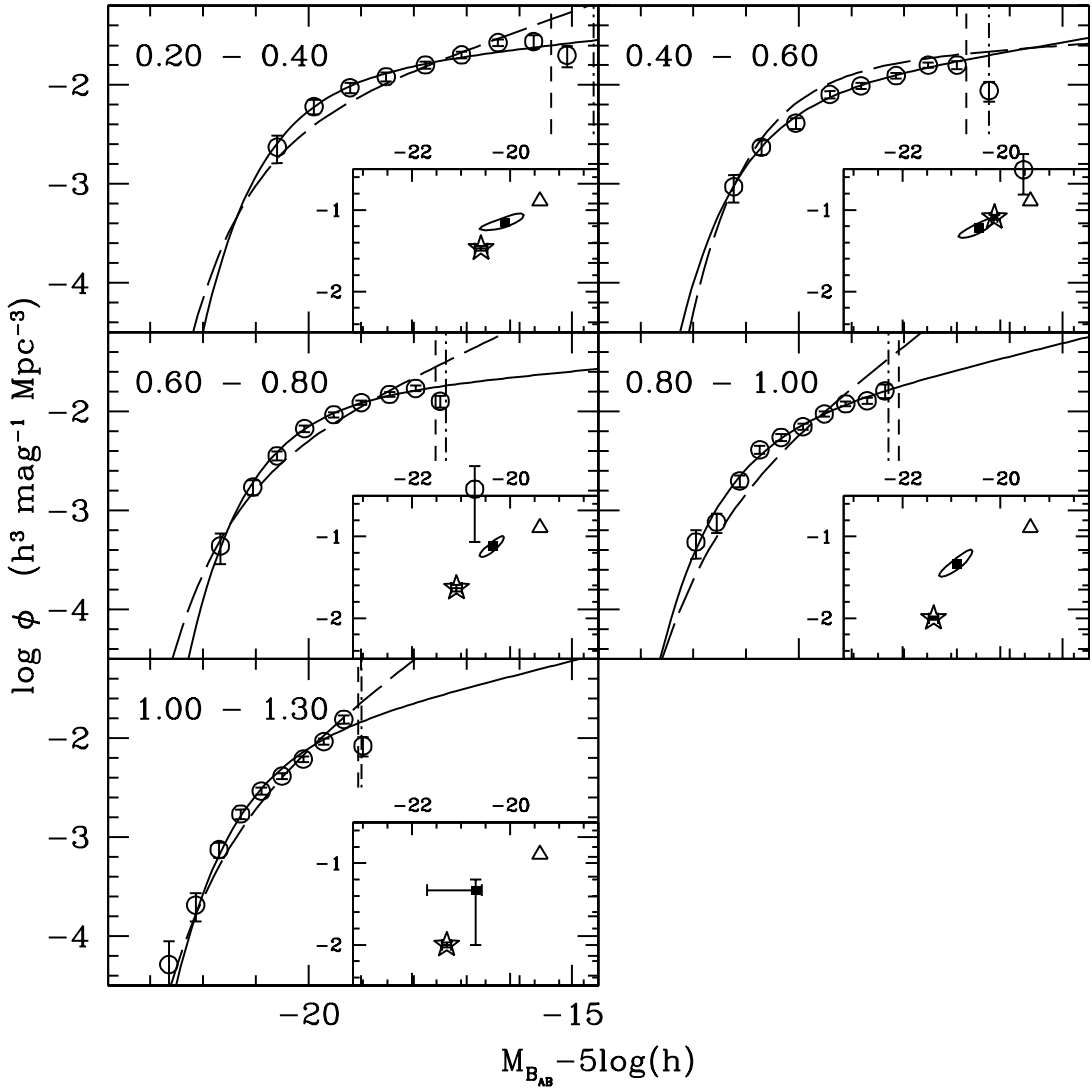}
\caption{ Comparison between the COMBO-17 and 
the VVDS global $B$-band LFs. The solid lines and the points
correspond to the VVDS estimates. The vertical short-dashed lines are
the faint absolute magnitude limits considered in the STY
estimates. The long dashed lines are the global LFs derived by
\cite{Wolf03}(2003).  The vertical dot-dashed lines
correspond to the faint absolute magnitude limits surveyed by the
COMBO-17 data. The best estimated values for the
$\alpha,M^*$ parameters measured by the VVDS are shown with solid
squares in insets, and the associated 90\% error contours with solid
lines. The ($\alpha, M^*$) parameters given by COMBO-17 are the open
stars, with error bars included inside the symbols. }
\label{figureCombo}
\end{center}
\end{figure*}

Wolf et al. (2003) have derived the LFs up to $z=1.2$ with
a sample of $\sim$ 25,000 galaxies from the COMBO-17
survey. This sample is selected in the $R$ band ($R_{vega} \le
24$). The redshifts are photometric redshifts derived from medium-band
photometry in 17 filters. The Schechter parameters of the COMBO-17
global LF are available in the online material of the paper
(\cite{Wolf03}2003). The comparison between the $B$-band global LFs of
VVDS and COMBO-17 surveys is shown in the Fig.\ref{figureCombo} in
five redshift bins up to z = 1.2.

The bright parts of the LFs appear to be roughly in agreement,
although some significant differences are seen in a few redshift bins
(see, for example, the redshift bins [0.4-0.6] and [0.8-1]). Given the
errors on the $\alpha-M^*$ parameters reported by the two surveys, the
overall LF shapes are not consistent with each other (see insets in
Fig.\ref{figureCombo}). Since the COMBO-17 sample is selected from the
$R$ band, its global $B$-band LF could be affected by the bias
described in \cite{Ilbert04}(2004) at $z > 0.5$. This bias
introduces an overestimate of the LF faint-end at $z > 0.5$ and could
explain the significantly steeper slope measured by the COMBO-17
survey in the redshift bins [0.6-0.8] and [0.8-1]. Since $\alpha$ and
$M^*$ are correlated, the same effect could also explain the
differences seen in the bright part of the LF. These discrepancies can
also be due to other reasons as, for example, a smaller fraction of
very blue galaxies in the I-selected VVDS sample (in fact,
the LF of the bluest galaxies has the steepest faint-end
LF) or a bias in the COMBO-17 estimate due to their use of photometric
redshifts. These possibilities will be better investigated through a
comparison of the COMBO-17 and VVDS LFs for each galaxy
type (\cite{Zucca05}2005), since such a comparison is much less
affected by the bias discussed above (\cite{Ilbert04}2004).

\section{Conclusions}

We use the first epoch spectroscopic deep sample of the VVDS, with
11,034 spectra selected up to $I_{AB}=24$, to derive the global LF up
to $z=2$ in the five bands $U$, $B$, $V$, $R$, $I$. The global LFs are
measured using ALF and care is taken to remove the bias introduced by
the difference of visibility of the different galaxy spectral types.

We observe a clear evolution of the global LF with redshift in all
bands and we find that this evolution is significantly dependent on
the rest-frame wavelength, being stronger at shorter wavelengths.  The
comoving density of the bright galaxies increases with redshift from
$z = 0.05$ up to $z = 1$. This increase is by a factor $\sim 2.6$ in
the $U$ band and becomes smaller for redder rest-frame wavelengths,
with values of the order of $2.6, 2.2, 1.8, 1.5, 1.5$ in the $U$, $B$,
$V$, $R$, $I$ bands, respectively.

In order to better distinguish the processes responsible 
of this evolution, we have studied the evolution with 
redshift of the Schechter parameters computed with the STY
estimator. This analysis suggests a possible steepening of the slope
with redshift. The observed change in $\alpha$ is $\sim -0.3$ from
$z=0.05$ up to $z=1$, similar in all bands. This evolution
has to be confirmed with the on-going second epoch VVDS 
data, which will allow us to decrease significantly the statistical
errors on $\alpha$. This evolution of the global LF slope is 
expected because of the different evolutions observed
for the different galaxy types (\cite{Zucca05}2005). In particular,
since the LF of blue galaxies has a steep slope and evolves strongly
with redshift (e.g., \cite{Lilly95}1995, \cite{Zucca05}2005), the
relative contribution of the blue population to the global LF
increases with redshift and could explain the steepening
of the slope.

We also measure a significant brightening of the global LF with
redshift. This brightening, parameterized as the change of the best
fit value of $M^*$, is a function of the rest-frame wavelength.
Compared to the local SDSS values, we obtain a brightening included in
the range $1.8-2.5$, $1.7-2.4$, $1.2-1.9$, $1.1-1.8$ and $1.0-1.6$
magnitudes from $z = 0.05$ up to $z = 2$, in the $U$, $B$,
$V$, $R$ and $I$ rest-frame bands. This tendency of a stronger
brightening toward bluer rest-frame wavelengths is consistent with
existing measurements at shorter and longer rest-frame wavelengths.
In the rest-frame far-UV (1530 \AA), \cite{Arnouts05} 
measure a brightening $\Delta M^* \sim -2$ magnitudes up to $z = 1$,
stronger than our measurement in the $U$ band in the same redshift
interval. In the near-IR, \cite{Pozzetti03} measure an
evolution consistent with a mild luminosity evolution both in the $J$
and $K$ bands with $\Delta M_J \sim -0.7$ and $\Delta M_K \sim -0.5$
at $z \sim 1$. This differential evolution of $M^*$ with wavelength is
expected, since the rest-frame luminosity at different wavelengths
probes different physical processes acting in galaxy formation and
evolution. The fact that the brightening is stronger in the bluest
bands suggests that most of the evolution of the global LFs up to $z =
2$ is related to the star formation history, better probed with the
luminosity measured at short rest-frame wavelengths.  The luminosity
density and star formation rate derived from the VVDS first epoch
observations will be presented in \cite{Tresse05}(2005).  We will
explore the evolution of the LFs per spectral types and as a function
of the local environment in forthcoming papers (\cite{Zucca05}2005,
\cite{Ilbert05}2005).

\begin{table*}
\begin{center}
\begin{tabular}{c c c c c c c} \hline

\multicolumn{7}{c}{$\Omega_m$=0.3 \hspace{1cm}                $\Omega_\Lambda$=0.7} \\ \hline
  &  &  & & & & \multicolumn{1}{c}{$\Phi^*$}      \\
Band & z-bin &  Number$^{(a)}$ &  Number$^{(b)}$ &   $\alpha$ &  $M^*_{AB}-5log(h)$ & ($10^{-3} h^3 Mpc^{-3}$)                    \vspace{0.2cm} \\ \hline
\hline
U    &    0.05-0.20 &   233 &   205 &  -1.05$^{{\rm + 0.05}}_{{\rm -0.05}}$ & -18.18 &  26.43$^{{\rm + 3.91}}_{{\rm -3.53}}$ \\ 
     &    0.20-0.40 &   928 &   728 &  -1.17$^{{\rm + 0.07}}_{{\rm -0.07}}$ & -18.83$^{{\rm + 0.17}}_{{\rm -0.19}}$ &  18.16$^{{\rm + 3.60}}_{{\rm -3.30}}$ \\ 
     &    0.40-0.60 &  1250 &   888 &  -1.17$^{{\rm + 0.09}}_{{\rm -0.09}}$ & -19.43$^{{\rm + 0.16}}_{{\rm -0.18}}$ &  13.39$^{{\rm + 2.67}}_{{\rm -2.51}}$ \\ 
     &    0.60-0.80 &  1793 &  1057 &  -1.27$^{{\rm + 0.12}}_{{\rm -0.12}}$ & -19.52$^{{\rm + 0.15}}_{{\rm -0.17}}$ &  14.52$^{{\rm + 3.08}}_{{\rm -2.89}}$ \\ 
     &    0.80-1.00 &  1508 &   935 &  -1.44$^{{\rm + 0.20}}_{{\rm -0.19}}$ & -19.75$^{{\rm + 0.22}}_{{\rm -0.26}}$ &  12.43$^{{\rm + 4.40}}_{{\rm -4.07}}$ \\ 
     &    1.00-1.30 &  1200 &   902 &  -1.44 & -19.89$^{{\rm + 0.21}}_{{\rm -0.70}}$ &  10.32$^{{\rm + 3.49}}_{{\rm -7.27}}$ \\ 
     &    1.30-2.00 &   477 &   468 &  -1.44 & -20.17$^{{\rm + 0.17}}_{{\rm -0.51}}$ &   5.14$^{{\rm + 1.26}}_{{\rm -3.05}}$ \\ 
\hline
B    &    0.05-0.20 &   233 &   227 &  -1.09$^{{\rm + 0.05}}_{{\rm -0.05}}$ & -19.39 &  21.19$^{{\rm + 3.20}}_{{\rm -2.88}}$ \\ 
     &    0.20-0.40 &   928 &   891 &  -1.15$^{{\rm + 0.05}}_{{\rm -0.05}}$ & -20.09$^{{\rm + 0.19}}_{{\rm -0.22}}$ &  14.60$^{{\rm + 2.53}}_{{\rm -2.37}}$ \\ 
     &    0.40-0.60 &  1250 &  1172 &  -1.22$^{{\rm + 0.06}}_{{\rm -0.06}}$ & -20.45$^{{\rm + 0.16}}_{{\rm -0.18}}$ &   9.62$^{{\rm + 1.68}}_{{\rm -1.56}}$ \\ 
     &    0.60-0.80 &  1793 &  1724 &  -1.12$^{{\rm + 0.06}}_{{\rm -0.06}}$ & -20.36$^{{\rm + 0.11}}_{{\rm -0.12}}$ &  15.07$^{{\rm + 1.93}}_{{\rm -1.86}}$ \\ 
     &    0.80-1.00 &  1508 &  1507 &  -1.33$^{{\rm + 0.08}}_{{\rm -0.08}}$ & -20.87$^{{\rm + 0.15}}_{{\rm -0.17}}$ &   9.07$^{{\rm + 1.78}}_{{\rm -1.67}}$ \\ 
     &    1.00-1.30 &  1200 &  1177 &  -1.33 & -20.70$^{{\rm + 0.13}}_{{\rm -0.99}}$ &  11.62$^{{\rm + 2.26}}_{{\rm -9.46}}$ \\ 
     &    1.30-2.00 &   477 &   382 &  -1.33 & -21.20$^{{\rm + 0.09}}_{{\rm -0.62}}$ &   4.31$^{{\rm + 0.52}}_{{\rm -2.80}}$ \\ 
\hline
V    &    0.05-0.20 &   233 &   231 &  -1.15$^{{\rm + 0.04}}_{{\rm -0.04}}$ & -20.44 &  14.75$^{{\rm + 2.60}}_{{\rm -2.29}}$ \\ 
     &    0.20-0.40 &   928 &   921 &  -1.21$^{{\rm + 0.04}}_{{\rm -0.04}}$ & -20.89$^{{\rm + 0.23}}_{{\rm -0.28}}$ &  10.46$^{{\rm + 1.98}}_{{\rm -1.85}}$ \\ 
     &    0.40-0.60 &  1250 &  1250 &  -1.35$^{{\rm + 0.05}}_{{\rm -0.05}}$ & -21.56$^{{\rm + 0.22}}_{{\rm -0.27}}$ &   5.17$^{{\rm + 1.14}}_{{\rm -1.05}}$ \\ 
     &    0.60-0.80 &  1793 &  1780 &  -1.35$^{{\rm + 0.05}}_{{\rm -0.05}}$ & -21.38$^{{\rm + 0.14}}_{{\rm -0.15}}$ &   7.33$^{{\rm + 1.20}}_{{\rm -1.12}}$ \\ 
     &    0.80-1.00 &  1508 &  1365 &  -1.50$^{{\rm + 0.07}}_{{\rm -0.07}}$ & -21.85$^{{\rm + 0.19}}_{{\rm -0.22}}$ &   4.42$^{{\rm + 1.16}}_{{\rm -1.04}}$ \\ 
     &    1.00-1.30 &  1200 &   969 &  -1.50 & -21.57$^{{\rm + 0.33}}_{{\rm -0.85}}$ &   6.12$^{{\rm + 3.82}}_{{\rm -4.71}}$ \\ 
     &    1.30-2.00 &   477 &   273 &  -1.50 & -21.86$^{{\rm + 0.21}}_{{\rm -0.48}}$ &   2.89$^{{\rm + 0.95}}_{{\rm -1.65}}$ \\ 
\hline
R    &    0.05-0.20 &   233 &   233 &  -1.16$^{{\rm + 0.04}}_{{\rm -0.04}}$ & -20.82 &  13.71$^{{\rm + 2.45}}_{{\rm -2.14}}$ \\ 
     &    0.20-0.40 &   928 &   928 &  -1.27$^{{\rm + 0.04}}_{{\rm -0.04}}$ & -21.64$^{{\rm + 0.30}}_{{\rm -0.41}}$ &   7.19$^{{\rm + 1.66}}_{{\rm -1.58}}$ \\ 
     &    0.40-0.60 &  1250 &  1244 &  -1.42$^{{\rm + 0.04}}_{{\rm -0.04}}$ & -22.20$^{{\rm + 0.27}}_{{\rm -0.35}}$ &   3.40$^{{\rm + 0.89}}_{{\rm -0.83}}$ \\ 
     &    0.60-0.80 &  1793 &  1685 &  -1.41$^{{\rm + 0.05}}_{{\rm -0.05}}$ & -21.85$^{{\rm + 0.15}}_{{\rm -0.17}}$ &   5.55$^{{\rm + 1.01}}_{{\rm -0.93}}$ \\ 
     &    0.80-1.00 &  1508 &  1214 &  -1.53$^{{\rm + 0.08}}_{{\rm -0.07}}$ & -22.31$^{{\rm + 0.21}}_{{\rm -0.25}}$ &   3.41$^{{\rm + 1.00}}_{{\rm -0.89}}$ \\ 
     &    1.00-1.30 &  1200 &   841 &  -1.53 & -21.99$^{{\rm + 0.38}}_{{\rm -0.84}}$ &   4.68$^{{\rm + 3.48}}_{{\rm -3.58}}$ \\ 
     &    1.30-2.00 &   477 &   220 &  -1.53 & -22.17$^{{\rm + 0.24}}_{{\rm -0.46}}$ &   2.49$^{{\rm + 0.93}}_{{\rm -1.38}}$ \\ 
\hline
I    &    0.05-0.20 &   233 &   233 &  -1.19$^{{\rm + 0.04}}_{{\rm -0.04}}$ & -21.18 &  11.80$^{{\rm + 2.22}}_{{\rm -1.92}}$ \\ 
     &    0.20-0.40 &   928 &   928 &  -1.32$^{{\rm + 0.04}}_{{\rm -0.04}}$ & -22.46$^{{\rm + 0.45}}_{{\rm -0.75}}$ &   4.73$^{{\rm + 1.45}}_{{\rm -1.53}}$ \\ 
     &    0.40-0.60 &  1250 &  1220 &  -1.47$^{{\rm + 0.04}}_{{\rm -0.04}}$ & -22.75$^{{\rm + 0.33}}_{{\rm -0.45}}$ &   2.44$^{{\rm + 0.76}}_{{\rm -0.71}}$ \\ 
     &    0.60-0.80 &  1793 &  1576 &  -1.41$^{{\rm + 0.05}}_{{\rm -0.05}}$ & -22.17$^{{\rm + 0.16}}_{{\rm -0.18}}$ &   5.01$^{{\rm + 0.94}}_{{\rm -0.86}}$ \\ 
     &    0.80-1.00 &  1508 &  1101 &  -1.52$^{{\rm + 0.08}}_{{\rm -0.08}}$ & -22.63$^{{\rm + 0.22}}_{{\rm -0.26}}$ &   3.07$^{{\rm + 0.94}}_{{\rm -0.83}}$ \\ 
     &    1.00-1.30 &  1200 &   748 &  -1.52 & -22.32$^{{\rm + 0.37}}_{{\rm -0.89}}$ &   4.02$^{{\rm + 2.92}}_{{\rm -3.14}}$ \\ 
     &    1.30-2.00 &   477 &   189 &  -1.52 & -22.38$^{{\rm + 0.22}}_{{\rm -0.44}}$ &   2.51$^{{\rm + 0.84}}_{{\rm -1.34}}$ \\ 
\hline
\multicolumn{7}{l}{(a) Number of galaxies in the  redshift bin (sample used for 1/V$_{\rm max}$ estimate)} \\
\multicolumn{7}{l}{(b) Number of galaxies brighter than bias limit (sample used for STY, C+, SWML estimate)} \\

\end{tabular}
\caption{Schechter parameters and associated one sigma errors
  ($2\Delta ln\mathcal L=1$) of the global LFs between $z=0.05$ and
  $z=2$ and derived in the $U$, $B$, $V$, $R$, and $I$ filters of the
  standard system. Parameters listed without errors are set `ad-hoc'
  to the given value.}
\label{table1}
\end{center}
\end{table*}

\section*{Acknowledgments}

This research has been developed within the framework of the VVDS
consortium.\\
We thank the ESO staff at Paranal for their help in the acquisition of
the data. We thank C. Moreau at LAM for the installation
of our code, ALF, under the VVDS Database.\\
This work has been partially supported by the CNRS-INSU and its
Programme National de Cosmologie (France) and Programme National
Galaxies (France), and by Italian Ministry (MIUR) grants
COFIN2000 (MM02037133) and COFIN2003 (num.2003020150).\\
The VLT-VIMOS observations have been carried out on guaranteed time
(GTO) allocated by the European Southern Observatory (ESO) to the
VIRMOS consortium, under a contractual agreement between the Centre
National de la Recherche Scientifique of France, heading a consortium
of French and Italian institutes, and ESO, to design, manufacture and
test the VIMOS instrument.

\appendix

\section{The Algorithm for Luminosity Function (ALF)}

This section describes the standard methods implemented in our
Algorithm for Luminosity Function (ALF) developed within the VVDS
framework. We present how we derive the rest-frame absolute magnitudes
and the details of the 1/Vmax, C$^+$, SWML and STY estimators
implemented in this tool.

\subsection{Absolute magnitudes}

\begin{figure}
  \centering 
\includegraphics[width=9cm]{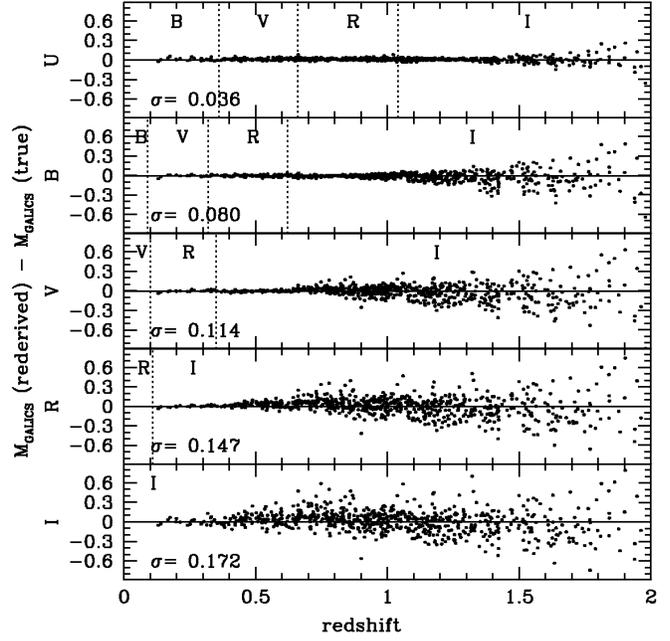}
\caption{Difference between our `rederived' absolute magnitudes  
  and `true' absolute magnitudes from GALICS as a function of
  redshift, in the $U$, $B$, $V$, $R$ and $I$ reference bands from
  the top to the bottom panels, respectively. The vertical lines
  indicate the change of $Obs$ filter and the adopted $Obs$ filter is
  labeled in the corresponding redshift range.}
\label{figureMag}
\end{figure}

The k-correction depends on the galaxy spectral energy distribution
(SED). At high redshift, it is the main source of error and systematic
in the absolute magnitude measurement. Using {\it Le Phare}, we adjust
the best SED template on $U$, $B$, $V$, $R$ and $I$ apparent magnitudes to
derive k-corrections.  We use a set of templates generated
with the galaxy evolution model PEGASE.2 (\cite{Fioc97}).  The
templates are computed for eight spectral classes
including elliptical, spiral, irregular and starburst, with the
initial mass function (IMF) from \cite{Rana92}, with ages 
varying between 10 Myr and 14 Gyr.  Dust extinction and metal effects
are included, depending on the evolution scenario.  We derive the
absolute magnitude in the reference band $Ref$ from the apparent
magnitude in the band $Obs$:
\begin{equation}
    M^{Ref}=m^{Obs}-DM(z,H_0,\Omega_m,\Omega_\Lambda)-KC(z,{\rm SED})
\label{eqn:Magabs}
\end{equation}
with DM the distance modulus and $KC$ defined as follows:
\begin{equation}
KC(z, {\rm SED}) = (k^{Ref}(z) + m^{Obs}(z) - m^{Ref}(z))^{{\rm SED}}
\label{eqn:KC}
\end{equation}
with $k^{Ref}$ the k-correction in the reference band (\cite{Oke68}).
To limit the template dependency, the $Obs$ band is chosen
automatically to be the closest as possible to the $Ref$
band redshifted in the observer frame. We do not correct the absolute
magnitudes for the internal dust extinction related to the considered
galaxies.

To check the robustness of our absolute magnitude estimate, we use the
GALICS simulations (\cite{Hatton03}). We extract a simulated catalogue
with $B$, $V$, $R$, $I$ apparent magnitudes and redshifts from the
GALICS/MOMAF database. We apply exactly the method described before,
to rederive the absolute magnitudes.  Fig.\ref{figureMag} shows the
difference between our measurements and the `true' absolute magnitudes
from GALICS. When our procedure to limit the template
dependency is efficient, the dispersion remains very small in
comparison to the photometric errors. For instance, $\sigma \sim
0.03-0.04$ in the U-band, and we limit the template dependency up to
$z=2$. If our procedure can not be applied (since NIR data are not
considered here), the dispersion increases. For instance, in the
$I$-band the dispersion due to the k-correction is $\sim 0.2$.

\subsection{Luminosity function estimators}
  
We describe in this subsection the four standard estimators
implemented in ALF, the 1/V$_{\rm max}$, C$^{+}$, SWML and STY
estimators.  

\subsubsection{The 1/V$_{\rm max}$ estimator}
     
The 1/V$_{\rm max}$ LF estimator (\cite{Schmidt68}) is the most often
used to derive the LF, because of its simplicity. This estimator
requires no assumption on the luminosity distribution. The 1/V$_{\rm
max}$ gives directly the normalization of the LF, assuming implicitly
an uniform spatial distribution of the
galaxies.

We consider a sample selected between bright and faint apparent
magnitude limits, $m_b$ and $m_f$ respectively. The maximum
observable comoving volume in which galaxy~$i$ can be detected is
given by
\begin{equation} \label{eqn:Vol}
V_{obs, i} = \int_\omega\int^{z_{max, i}}_{z_{min, i}}\frac{d^2V} {d\omega dz} d\omega dz, 
\end{equation}
where $\omega$ is the effective solid angle of the survey, and $V$ is
the comoving volume. $z_{min, i}$ and $z_{max, i}$ are the lower and
upper redshift limits within a galaxy~$i$ can be included in the
sample. The LF, $\phi(M)$, is discretized in absolute magnitudes
\begin{equation} \label{eqn:rebinVmax}
\phi(M) = \sum^{N_{bin}}_{k=1} \phi_k W(M_k - M),
\end{equation}
where the window function $W$ is defined as, 
\begin{equation} \label{eqn:W}
W(x) = 
\left\{
\begin{array}{rl}
1 & \mbox{if $-dM/2 \le x < dM/2 $} \\
0 & \mbox{otherwise.}
\end{array}
\right.
\end{equation}
$\phi_k$ is derived in each absolute magnitude bin $k$ as follows:
\begin{equation} \label{eqn:PhiVmax}
\phi_k dM = \sum^{N_g}_{i=1}\frac{w_i}{V_{obs, i}}W(M_k - M_i),
\end{equation}
where $N_g$ is the number of galaxies of the sample and $w_i$ is the
weight applied to correct the unidentified sources in the field (see
Section~3). We associate the Poisson errors to the 1/V$_{\rm max}$
following (\cite{Marshall85}):
\begin{equation}\label{eqn:Vmaxerror}
\sigma_{\phi_k}=\sqrt{\sum_{i=1}^{N_g}W(M_k - M_i)\frac{w_i^2}{{V_{{obs}_i}}^2}}.
\end{equation}

\subsubsection{The C$^+$ estimator}
     
\noindent   \cite{Lynden-Bell71}(1971) derived the non-parametric 
C$^-$ method to overcome the assumption of a uniform galaxy
distribution. We use a modified version of the C$^-$, called
C$^+$ (\cite{Zucca97}). This method is based on the
equality:
\begin{equation}\label{eqn:lynden}
\frac{d\psi}{\psi}=\frac{dX}{C^+},
\end{equation}
with $\psi$ the cumulative luminosity function, $d\psi$ the variation
of the cumulative luminosity function between $M$ and $M+dM$. $dX$ is
the number of observed galaxies between $M$ and $M+dM$ and $C^+$ is
the number of galaxies brighter than $M+dM$ with a redshift lower than the
maximum redshift observable. We use a sample sorted from the faintest
to the brightest absolute magnitude. We note $C^+_i$ the value of
$C^+$ for a galaxy $i$. We introduce the weight $w_i$ in $C^+_i$ as
follows:
\begin{equation}\label{eqn:C+}
C^+_i=\sum^{N_g}_{j=i , z_j \le z_{max, i}}w_j.
\end{equation}
We note $f_i$ the variation of the cumulative luminosity function in
the neighborhood of the galaxy $i$, between $M_i^-$ and $M_i^+$. We
can write the cumulative luminosity function:
\begin{equation}\label{eqn:cum_f}
\psi(M_i)=\psi_i=\sum^{N_g}_{j=i}f_j=\psi_0-\sum_{j=1}^{i-1}f_j.
\end{equation}
We impose the limit values $f_0=0$ and $\psi_0=1$ to normalize the
cumulative luminosity function. We obtain the recurrence relation,
used to derive the contribution of all galaxies in the sample:
\begin{equation}\label{eqn:recC+}
f_i=\frac{\psi_i}{C^+_i}= \frac{1-\sum_{j=1}^{i-1}f_j}{C^+_i}.
\end{equation}
The LF is given by rebinning the contributions $f_i$:
\begin{equation} \label{eqn:rebinC+}
\phi_k dM = \sum_{i=1}^{N_g} f_i w_i W(M_k - M_i). 
\end{equation}
Poisson errors are associated as done for the 1/V$_{\rm
max}$ estimator (Eq.~\ref{eqn:Vmaxerror}).

\subsubsection{The STY and SWML estimators}

The STY (\cite{Sandage79}) and the SWML
(\cite{Efstathiou88}, hereafter EEP88) estimators are both
derived from maximum likelihood method. The likelihood $\mathcal L$ is
the joint probability of observing the galaxy sample, taking into account the
observational selection effects. The principle of the SWML and STY is
to maximize $\mathcal L$ with respect to the LF. $\mathcal L$ is given
by:
\begin{equation} \label{eqn:LikelihoodDef} 
\mathcal L = \prod^{N_g}_{i=1} p(M_i,z_i)^\frac{w_i}{<w>} = \prod^{N_g}_{i=1}\left(\frac{\phi(M_i)}{\int^{M_{f, i}}_{M_{b, i}} \phi(M)dM}\right)^\frac{w_i}{<w>}
\end{equation}
where $M_{f, i}$ and $M_{b, i}$ are the faint and bright observable
absolute magnitudes of a galaxy~$i$ at redshift $z_i$. The weight is
introduced in $\mathcal L$ following \cite{Zucca94}.  This weight
artificially decreases the size of the error contours derived from the
analysis of $\mathcal L$, then we balance this weight by the average
weight $<w>$. The average weight $<w>$ does not
affect the minimization of $\mathcal L$.\\

The STY assumes a functional form for the luminosity distribution. We
use the empirical Schechter function (\cite{Schechter76}):
\begin{equation} \label{eqn:schechter}
\phi(L)dL=\phi^*e^{-\frac{L}{L^*}}\left(\frac{L}{L^*}\right)^{\alpha}d\left(\frac{L}{L^*}\right).
\end{equation}
The likelihood (eq.\ref{eqn:LikelihoodDef}) can be written 
as:
\begin{equation}\label{eqn:L_STY}
\begin{array}{lll}
ln\mathcal L &=&\frac{1}{<w>}[\alpha\sum^{N_g}_{i=1}w_i ln L_i-\\
             & &(1/L^*)\sum^{N_g}_{i=1}w_i L_i-(\alpha+1)lnL^*\sum^{N_g}_{i=1}w_i \\
                  & &-\sum^{N_g}_{i=1}w_i ln(\Gamma(\alpha+1,\frac{L_{b,i}}{L^*})-\Gamma(\alpha+1,\frac{L_{f,i}}{L^*}))]
\end{array}
\end{equation}
with $\Gamma$ the incomplete Euler gamma function. We use the MINUIT
package of the CERN library (\cite{James95}) to minimize $-2ln\mathcal
L$ (MIGRAD procedure), to obtain the non-parabolic error for each
parameter (MINOS procedure) and the error contour $\alpha-M^*$ (MNCONT
procedure). The crosses of the likelihood surface with $ln\mathcal
L_{max}-\Delta ln\mathcal L$ is used to compute the errors. The
threshold $\Delta ln\mathcal L$ is chosen in a standard way that
depends on the desired confidence level in the estimate (
e.g., $2\Delta ln\mathcal L=2.3$ and $2\Delta ln\mathcal L=4.61$ to
estimate the $\alpha-M^*$ error contours with 68\% and 90\% confidence
level ; $2\Delta ln\mathcal L=1$ to estimate the one sigma error for
one parameter).\\

The SWML does not assume any functional form for the luminosity
distribution. The LF is discretized in absolute magnitude bins like
the 1/V$_{\rm max}$ (see Eq.~\ref{eqn:rebinVmax}). We maximize
$ln\mathcal L$ with respect to $\phi_k$ to obtain the recurrence
equation:
\begin{equation} \label{eqn:recurenceSWML}
\phi_j dM=\frac{\sum^{N_g}_{i=1}w_i W(M_i-M_j)}{\sum^{N_g}_{i=1}\frac{w_i H(M_{b,i}-M_j)H(M_j-M_{f,i})}{\sum^{N_{bin}}_{k=1}\phi_k dM H(M_{b,i}-M_k)H(M_k-M_{f,i})}}.
\end{equation}
with
\begin{equation} \label{eqn:W}
H(x) = 
\left\{
\begin{array}{rl}
0 & \mbox{if $x \le -dM/2 $} \\
x/dM+1/2 & \mbox{if $-dM/2 \le x \le dM/2 $} \\
1 & \mbox{otherwise.}
\end{array}
\right.
\end{equation}

We add a constraint $g$ on $\phi_k$ and rewrite the likelihood as
$ln\mathcal L'=ln\mathcal L+\lambda g$ where $\lambda$ is a Lagrangian
multiplier. Following EEP88, we choose $g=\sum^{N_{bin}}_{k=1}\phi_k
dM (L_k/L_f)^\beta-1$ with $L_f$ the fiducial luminosity and $\beta$ a
constant. The error bars are derived from the covariance matrix,
denoted $C$, defined as the inverse of the information matrix $I$:
\begin{equation} \label{eqn:cov_matrix}
C(\phi_k)=I^{-1}(\phi_k)=-\left(
\begin{array}{cc}
\frac{\delta^2 ln \mathcal L}{\delta \phi_i \delta \phi_j}+ 
\frac{\delta
g}{\delta \phi_i}\frac{\delta g}{\delta \phi_j} & \frac{\delta g}{\delta
\phi_j} \\ \frac{\delta g}{\delta \phi_i} & 0
\end{array}
\right)_{\phi_k}^{-1}.
\end{equation}
The second derivative of the likelihood is given by:
\begin{equation} \label{eqn:second_der_L}
\begin{array}{lcl}
\frac{\delta^2 ln \mathcal L}{\delta \phi_i \delta \phi_j} & = &
-\frac{1}{<w>}\sum^{N_g}_{l=1}\frac{w_l\delta_{ij}W(M_l-M_j)dM^2} 
{( \phi_jdM /(g+1))^2}+ \\
& &\frac{1}{<w>}\sum^{N_g}_{l=1}\frac{w_l dM^2 H1}{(\sum^{N_{bin}}_{k=1} \phi_k dM H2 /(g+1))^2}
\end{array}
\end{equation}
with $H1=H(M_{b,l}-M_i)H(M_i-M_{f,l})H(M_{b,l}-M_j)H(M_j-M_{f,l})$ and
 $H2=H(M_{b,l}-M_k)H(M_k-M_{f,l})$.
The error bars of the LF (for a normalization given by the constraint)
are given by the square root of the diagonal values of the covariance
matrix.

\subsubsection{Luminosity function normalization}
    
The estimators independent of the spatial density distribution (SWML,
STY and C$^+$) lose their normalization while the
normalization is directly done for the 1/V$_{\rm max}$ estimator. We
adopt the EEP88 density estimator to recover their normalization. The
density $n$ is simply the sum over all the galaxy sample of the
inverse of the selection function:
\begin{equation}\label{eqn:norma}
n=\frac{1}{V_{total}}\sum_{i=1}^{N_g}w_i\frac{\int^\infty_{-\infty}\phi(M)dM}{\int^{M_{f,i}}_{M_{b,i}}\phi(M)dM}.
\end{equation}
The comparison with the 1/V$_{\rm max}$ normalization is a direct and
independent check of the LF normalization.  The parameter $\phi^*$ is
directly related to the density $ \phi^*
\int^\infty_{-\infty}\phi(M)dM=n$.  $\phi^*$ is a function of $\alpha$
and $M^*$. To estimate the error on $\phi^*$, we derive $\phi^*$ for
the extreme values of the $\alpha-M^*$ error contour at one sigma
confidence level. We adopt Poisson errors when larger.

\end{document}